\documentclass[a4paper,fleqn]{cas-dc}

\usepackage[numbers,sort&compress]{natbib}

\usepackage{caption}
\usepackage{pifont} 

\usepackage{hyperref}
\hypersetup{colorlinks=true, citecolor=blue, linkcolor=blue, urlcolor=blue}

\usepackage{wrapfig} 
\usepackage{flushend}
\usepackage{appendix}

\def\tsc#1{\csdef{#1}{\textsc{\lowercase{#1}}\xspace}}
\tsc{WGM}
\tsc{QE}
\tsc{EP}
\tsc{PMS}
\tsc{BEC}
\tsc{DE}

\begin{document}
\let\WriteBookmarks\relax
\def\floatpagepagefraction{1}
\def\textpagefraction{.001}

\shorttitle{An Empirical Study of Excitation and Aggregation Design Adaptions in CLIP4Clip for Video-Text Retrieval}

\shortauthors{Xiaolun Jing et al.}

\title [mode = title]{An Empirical Study of Excitation and Aggregation Design Adaptions in CLIP4Clip for Video-Text Retrieval}                      
	
\author[1,2]{Xiaolun Jing}[orcid=0000-0002-1159-062X]

\ead{jingxiaolun@sjtu.edu.cn}

\credit{Conceptualization of this study, Methodology, Experiment}

\address[1]{Ningbo Artificial Intelligence Institute, Shanghai Jiao Tong University, China}
\address[2]{Department of Automation, Shanghai Jiao Tong University, China}

\author[1,2]{Genke Yang}[orcid=0000-0003-3492-0211]
\cormark[1]
\cortext[1]{Corresponding author}


\ead{gkyang@sjtu.edu.cn}

\author[1,2]{Jian Chu}

\ead{chujian@sjtu.edu.cn}

\begin{abstract}
CLIP4Clip model transferred from the CLIP has been the de-factor standard to solve the video clip retrieval task from frame-level input, triggering the surge of CLIP4Clip-based models in the video-text retrieval domain. 
In this work, we rethink the inherent limitation of widely-used mean pooling operation in the frame features aggregation and investigate the adaptions of excitation and aggregation design for discriminative video representation generation. We present a novel excitation-and-aggregation design, including (1) The excitation module is available for capturing non-mutually-exclusive relationships among frame features and achieving frame-wise features recalibration, and (2) The aggregation module is applied to learn exclusiveness used for frame representations aggregation. Similarly, we employ the cascade of sequential module and aggregation design to generate discriminative video representation in the sequential type. Besides, we adopt the excitation design in the tight type to obtain representative frame features for multi-modal interaction. The proposed modules are evaluated on three benchmark datasets of MSR-VTT, ActivityNet and DiDeMo, achieving MSR-VTT (43.9 R@1), ActivityNet (44.1 R@1) and DiDeMo (31.0 R@1). They outperform the CLIP4Clip results by +1.2\% (+0.5\%), +4.5\% (+1.9\%) and +9.5\% (+2.7\%) relative (absolute) improvements, demonstrating the superiority of our proposed excitation and aggregation designs. We hope our work will serve as an alternative for frame representations aggregation and facilitate future research.
\end{abstract}


\maketitle

\section{Introduction}
The past decade has witnessed the explosive growth of video media information. On the one hand, the advent of the big data era pushes the sharp increase in video quantity. On the other hand, an increasing number of videos are uploaded to various video media software (e.g. YouTuBe and TikTok) every day. Faced with countless videos of flexible lengths, an urgent problem arises: How can we accurately and efficiently search for the item to meet our requirements? One of the appropriate solutions is to exploit the semantics relevance between the language sentence and the video, thus successfully locating the needed video.

The video-text retrieval tasks are composed of four components: video representation extraction \cite{cheng2021improving,fang2021clip2video,han2021visual}, textual representation extraction \cite{zaremba2014recurrent,ging2020coot}, feature embedding and matching \cite{wang2022hybrid,jiang2022tencent,wang2022disentangled}, and objective functions \cite{gao2021clip2tv,ma2022x}. Among them, the enhancement of video representation extraction power has gained large popularity. Previously, CNNs-based methods have become the de-factor standard for video representation extraction in terms of spatial feature. Many popular CNNs variants like AlexNet \cite{krizhevsky2017imagenet}, VGGNet \cite{simonyan2014very}, ResNet \cite{he2016deep} and DenseNet \cite{huang2017densely}, are employed to extract the features from sampled video frames due to the superior characteristics of translation invariance and locality. For example, TCE \cite{yang2020tree} used a pre-trained ResNet-152 to extract video appearance features from the global average pooling. Likewise, NRCCR \cite{wang2022cross} extracted frame features from the FC layer whose input is the output of last average pooling layer. With the remarkable success of Transformer \cite{vaswani2017attention} in the natural language processing (NLP), a general Vision Transformer (ViT) \cite{dosovitskiy2020image} was provided for image or video frame features extraction, and it is widely applicable in the visual-language tasks, including image-text retrieval \cite{chen2019uniter, lee2018stacked}, visual entailment \cite{thomas2022fine, cao2022alignve} and visual question answering \cite{biten2022latr, cascante2022simvqa}. Among them, CLIP \cite{radford2021learning} utilized vision transformers to attain better performance within the given compute budget. GroupViT \cite{xu2022groupvit} adopted a hierarchical vision transformer to group semantic regions through language-driven mechanism. ViLD \cite{gu2021open} distilled the knowledge of pre-trained vision transformer into two-stage detector, achieving open-vocabulary object detection. Moreover, CLIP4Clip \cite{luo2022clip4clip} and ActionCLIP \cite{wang2021actionclip} employed  vision transformer encoder to extract video representation for video-text retrieval and video action classification tasks, bringing significant performance improvements. In addition to the single-modal video feature extraction, some works also study the extraction of multi-modal video features, such as appearance, motion and audio, each of which is extracted by a specific expert model. MoEE \cite{miech2018learning} was a such work, which utilized a mixture of embedding experts to learn text-video embeddings from heterogeneous data sources.

Recently, some works give priorities to the modeling of inherent noisy correspondence in the visual-text datasets to achieve further retrieval performance improvement. For instance, a Language Guidance (LG) model \cite{li2023integrating} was integrated into the Image-Text Matching (ITM) framework to identify and correct false negatives during training. CTPR \cite{feng2023learning} designed a novel partition criterion to divide training data into three subsets: hard, noisy, and clean pairs for the separate correspondence label smooth. In order to model intra-class variations in the vision-language data, Multi-View Visual Semantic Embedding (MV-VSE) \cite{li2022multi} was proposed to learn multiple embeddings for one visual data and improve the generalization ability of VSE methods. Furthermore, a Temporal Multi-modal Graph Transformer with Global-Local Alignment (TMMGT-GLA) \cite{feng2022temporal} was provided to model the input video as a sequence of semantic correlation graphs, thus effectively learning the cross-modal semantic correlations and temporal associations respectively.

\begin{figure*}[!t]
	\centerline{\includegraphics[width=\textwidth]{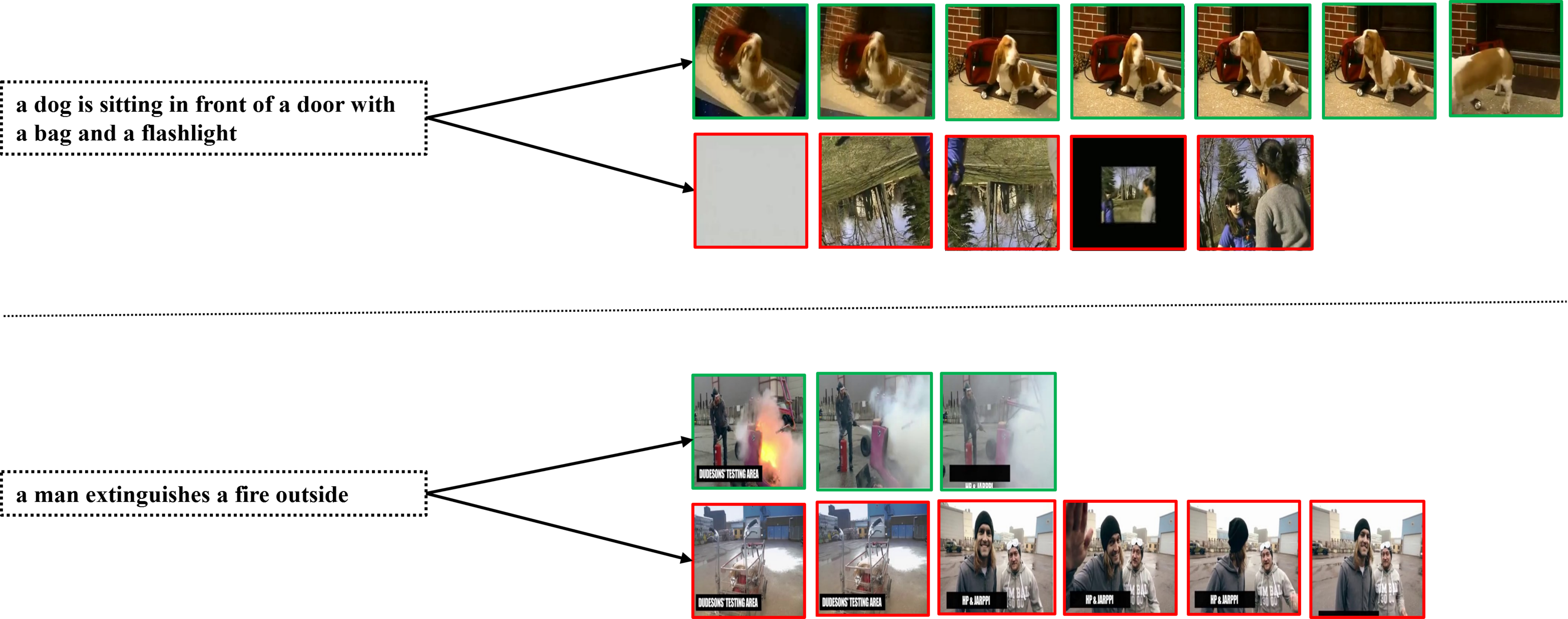}}
	\caption{Visualization of the semantic correlation discrepancies among distinct frames for single caption from the MSR-VTT dataset. Green boxes indicate that the caption and frame instances are semantic-relevant, while red boxes depict the semantic-irrelevant examples. Since not all frames are semantic-relevant to the given caption, aggregating frames features through mean pooling regardless of the frame content can be misleading. }
	\label{fig:motivation_visualization}
\end{figure*}

The proposed CLIP4Clip framework adopts a rough mean pooling method to aggregate frame features into video representation, which is used in the final similarity calculation. However, without the consideration of semantic correlation discrepancies among different caption-frame data, the obtained ``average frame'' is biased for cross-modal semantic matching. We also present the visualization results of semantic correlation discrepancies among multiple frames from the MSR-VTT \cite{xu2016msr} dataset in Fig. \ref{fig:motivation_visualization}. Here, we observe that distinct frames have diverse significances. In the top example, the caption describes the action of a dog, and surrounding objects, which are portrayed in the seven frames drawn in green boxes. Similarly, the flame extinction action expressed by the bottom caption is not semantic-aligned with the six frames drawn in red boxes. 

To address this issue, a Visual Semantic Enhanced Reasoning Network (ViSERN) \cite{feng2020exploiting} was devoted to viewing frame regions as vertices and the video is represented as a fully-connected semantic correlation graph, thus achieving the semantic reasoning between frame regions by random walk rule-based graph convolutional networks. Though proved to be effective, this method introduces a complex and time-consuming implementation to construct the semantic correlation graph from the input video, which hinders the widespread application in massive amount of video-text data scenarios. Therefore, our main effort is focused on the modeling of semantic correlation at frame-level features. More precisely, we transfer the method of Squeeze-and-Excitation module \cite{hu2018squeeze} to frame-wise relationship modeling and explore effective adaptions in different paradigms, including parameter-free, sequential, and tight types. Specifically, the excitation-and-aggregation module is substituted for mean pooling implementation to achieve frame-wise features aggregation. The hybrid module contains excitation and aggregation modules, of which the former models non-mutually-exclusive relationships among distinct frames, and the latter focuses on mutual exclusion modeling to generate representative video feature. In the sequential type, we utilize the combination of temporal modules like LSTM \cite{greff2016lstm}, and aggregation modules to aggregate sequential frame features into discriminative frame representation. Similarly, we propose the excitation module in the tight type, which is used for attentive frame enhancement and inattentive frame suppression before multi-modal interaction.

In summary, our contributions in this work are four-fold: 
(i) We propose a novel Excitation-and-Aggregation design to 
simultaneously capture non-mutually-exclusive and exclusive relationships among different frame representations by the usages of sigmoid and softmax activations in the excitation and aggregation stages; (ii) We present a new aggregation design to leverage the attention mechanism on the frame features output from the sequential modules like Transformer Encoder, and aggregate the sequential frame representations into discriminative video representation; (iii) We introduce an excitation design in the tight type to generate representative frame representations, which can achieve multi-modal interaction enhancement to understand text-video contents comprehensively;
(iv) We report the competitive retrieval performance on three text-video benchmarks and conduct extensive ablation experiments to demonstrate the effectiveness of our proposed approaches. 

The rest of this paper is organized as follows. Section \ref{Related Works} reviews the general excitation-relevant attention mechanisms utilization in the computer vision research, the development of video-text retrieval, and the recent proposed CLIP4Clip-based variants. Section \ref{Methodology} introduces the core components of CLIP4Clip, as well as the limitation of mean pooling operation and excitation and aggregation design adaptions in the improved similarity calculator. Section \ref{Experiment} gives extensive experimental results and detailed analysis, followed by the ablation experiments in section \ref{Ablation}.  Section \ref{Conclusion} presents the conclusions, limitations and future works of this empirical study. 

\section{Related Works}
\label{Related Works}
\subsection{Excitation-Relevant Attention Mechanisms}
As a milestone in the development of attention mechanism \cite{niu2021review,guo2022attention,fukui2019attention}, squeeze-and-Excitation network (SENet)
was devoted to adaptive channel-wise feature recalibration through modeling relationships among multiple channels. Based on the attention along channel dimension, CBAM \cite{woo2018cbam} further introduced spatial attention through the cascade of channel and spatial attention modules, achieving attention maps refinement along spatial and channel dimensions. To avoid dimensionality reduction in the SENet, ECA-Net \cite{wang2020eca} implemented 1D convolution to capture local cross-channel interaction within specified channel and adjacent channels. Besides, ESE-FN \cite{shu2022expansion} proposed expansion-squeeze-excitation attention mechanism to achieve multi-modal features fusion for elderly activity recognition. Unlike previous methods aiming at distinct channels or spatial regions in the input image, we mainly explore the effective adaptions of attention mechanism in video frame features aggregation. 

\subsection{Video-Text Retrieval}
Video-text retrieval is a non-trivial branch of multi-modal research, which aims to retrieval texts that are most semantic-relevant to a video-based search query. Most of the previous approaches tend to employ a single pre-trained language expert and a combination of task-specific or modality-specific pre-trained video experts for text and video features extraction. For example, CE \cite{liu2019use} proposed a collaborative experts model to aggregate different visual information, including object, scene, action, face, OCR, speech and audio features, into a compact video representation. MMT \cite{gabeur2020multi} designed a multi-modal transformer framework to jointly encode multiple modalities in  video, which is also leveraged to model the temporal relationship. However, the performance of these methods is largely limited by the pre-extracted single modal features, thus resulting in complex and biased retrieval. Moreover, some works benefited from end-to-end video-text retrieval paradigm, which directly trains own models on large-scale text-video datasets. Among them, ClipBERT \cite{lei2021less} presented a sparse sampling strategy to obtain clip-level predictions for end-to-end training, while Frozen in Time \cite{bain2021frozen} conducted end-to-end training on pairwise image-text and video-text datasets through uniformly sampling video frames. CLIP4Clip \cite{luo2022clip4clip} transferred the general image-text knowledge of CLIP \cite{radford2021learning} to video-text retrieval in an end-to-end manner, and investigated three similarity calculators for video-text contrastive learning. 

Although mean pooling implementation is widely used in the video feature generation, this rough aggregation scheme may inject irrelevant frame content into video representation and hinder the full utilization of 
consecutive video frames. We present excitation and aggregation modules to overcome this limitation and generate discriminative video feature for cross-modal semantic matching.

\subsection{CLIP4Clip-based Models}
Motivated by the outstanding performance of CLIP \cite{radford2021learning}, the work of CLIP4Clip \cite{luo2022clip4clip} extended the rich joint text-image understanding of CLIP \cite{radford2021learning} towards videos and boosted retrieval performance by a large margin (\textit{e.g.} more than 40\% relative improvement of the recall metric on MSR-VTT). Following the line of CLIP4Clip \cite{luo2022clip4clip}, some recent works were designed to obtain better retrieval performance. For instance, CLIP-ViP \cite{xue2022clip} proposed an Omnisource \cite{duan2020omni} Cross-modal Learning method equipped with a Video Proxy mechanism to tackle the domain gap between language sources. Clover \cite{huang2023clover} introduced a novel tri-modal alignment pre-training task to enhance cross-modal feature alignment and fusion. Additionally, UCO$\text{F}_{\text{I}}$A \cite{wang2023unified} was presented to consider cross-modal correspondence from different granularity, thus successfully capturing the multi-grained similarity between the text and video. HMMC \cite{shen2023end} adopted Hierarchical Matching and Momentum Contrast in an end-to-end pre-training network to explore the hierarchical semantic information in videos via multilevel semantic matching between videos and texts. However, these methods ignore semantic discrepancies among distinct video frames, bringing sub-optimal retrieval performance. Therefore, we propose excitation and aggregation modules to generate discriminative video representation, which is obtained through attentive frames enhancement and inattentive frames suppression.

\begin{figure*}[!t]
	\centerline{\includegraphics[width=\textwidth]{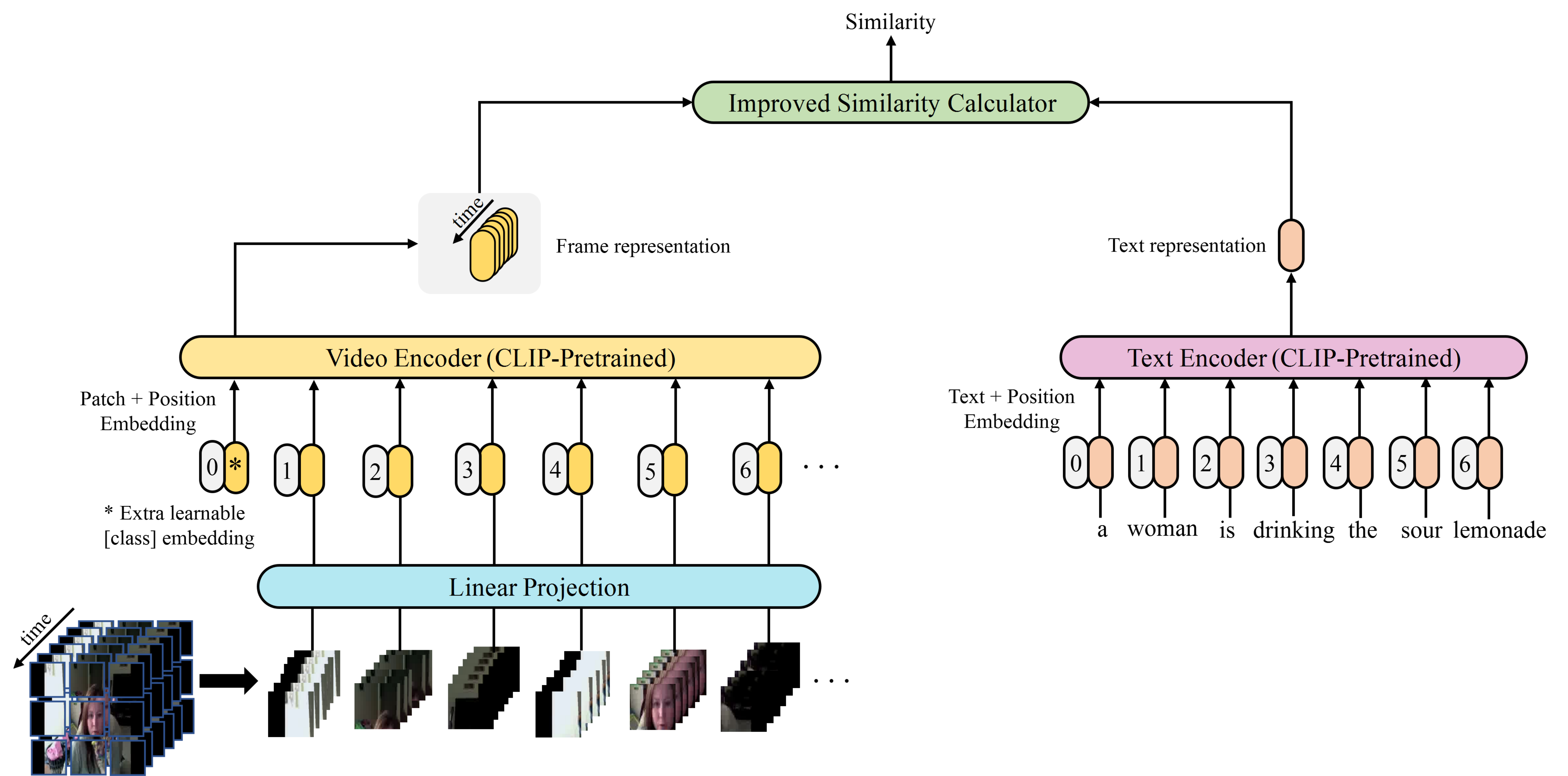}}
	\caption{The pipeline of our proposed method. The model integrates three core components, termed video encoder, text encoder and improved similarity calculator, of which the similarity score can be obtained from the output of improved similarity calculator. }
	\label{fig: main_structure}
\end{figure*}

\section{Methodology}
\label{Methodology}
This section presents the insights and methodologies that motivate our excitation and aggregation design adaptions in CLIP4Clip \cite{luo2022clip4clip}. Starting with a brief review of CLIP4Clip \cite{luo2022clip4clip} in Section \ref{sec: review}, we then explain the limitation of mean pooling implementation in frame-wise feature aggregation in Section \ref{sec: limitation}. Finally, we elaborate on the details of three improved similarity calculators in squeeze and expansion paradigms: (i) excitation-and-aggregation design (ii) aggregation design (iii) excitation design, in Section \ref{sec: similarity calculator}.

\subsection{A Brief Review of CLIP4Clip}
\label{sec: review}
CLIP4Clip \cite{luo2022clip4clip} is the first model to successfully transfer the knowledge of CLIP \cite{radford2021learning} to the text-video retrieval task, and it has achieved competitive performance in popular evaluation benchmarks such as MSR-VTT \cite{xu2016msr},  ActivityNet \cite{caba2015activitynet} and DiDeMo \cite{anne2017localizing}. The main idea of CLIP4Clip \cite{luo2022clip4clip} is to utilize the pre-trained CLIP \cite{radford2021learning} as a backbone to implement retrieval tasks between frame-level input and sentence-level input. The model is composed of three essential components: Video Encoder, Text Encoder, and Similarity Calculator.

\subsubsection{Video Encoder}
\label{sec: video}
The video encoder is ViT-B/32 with 12 layers and the patch size is 32. Most notably, the video encoder uses pre-trained CLIP \cite{radford2021learning} for better initialization to adapt the knowledge to the video domain. Given the input video frame, the ViT first divides the frame into non-overlapping image patches, then performs the 2D linear projection layers to map them into 1D tokens. After prepending a learnable [cls] token to the sequence of tokens and adding position embeddings to retain positional information, the tokens are fed into the transformer encoder to model interactions among patches. At the end, the [cls] token in the output from the highest layer plus layer normalization is treated as the frame representation. 

\subsubsection{Text Encoder}
\label{sec: text}
Similar to the video encoder, the text encoder is a transformer-based architecture, which is derived from the pre-trained 512-wide CLIP model, with 12 layers and 8 attention heads. CLIP4Clip \cite{luo2022clip4clip} regards the [EOS] token from the activations of last layer as generated caption representation. 

\subsubsection{Similarity Calculator}
\label{sec: sim_cal}
When the video frame-wise representation and the caption representation are obtained, CLIP4Clip \cite{luo2022clip4clip} designs similarity calculators in three categories depending on whether the module introduces extra learnable parameters. Specifically, the parameter-free type simply utilizes the mean pooling method to average frame features as video representation. Though proved to be effective, the mechanism is harmful to heavily temporal-dependent tasks. Therefore, the sequential type is introduced, and the cascade of general temporal modules like LSTM \cite{greff2016lstm}, Transformer encoder and the mean pooling method are utilized to generate a more powerful video representation. In addition, CLIP4Clip \cite{luo2022clip4clip} extends the modality of modeling information from frame-only to frame-caption level. The relationships between features of frames and caption are joint-modeling by the transformer encoder, and output at the first token from the last layer is fed into two linear projection layers plus an activation to compute the similarity score.

\begin{table}
	\caption{Semantic similarities between caption and multiple video frames at different frame lengths from the MSR-VTT dataset. ``Min Sim'' and ``Max Sim'' denote the minimal and maximal similarities respectively. ``MeanP Sim'' indicates the mean pooling similarity score.}
	\label{sim_comparison}
	\centering
	\resizebox{0.5\textwidth}{!}{
		\begin{tabular}{c|c|c|c|c}
			\hline 
			Frame Length & Video & Min Sim & Max Sim & MeanP Sim \\
			\hline 
			\multirow{4}*{12} & video3 & 0.2544 & 0.2996 & 0.2775 \\
			~ & video12 & 0.2310 & 0.2599 & 0.2444 \\
			~ & video28 & 0.1505 & 0.2409 & 0.2049 \\
			~ & video34 & 0.2302 & 0.2867 & 0.2503 \\
			\hline 
			\multirow{4}*{16} & video18 & 0.2241 & 0.3127 & 0.2756 \\
			~ & video52 & 0.2614 & 0.2961 & 0.2744 \\
			~ & video65 & 0.2678 & 0.3166 & 0.2905 \\
			~ & video96 & 0.2217 & 0.2915 & 0.2468 \\
			\hline
			\multirow{4}*{20} & video56 & 0.2529 & 0.3277 & 0.2869 \\
			~ & video165 & 0.2520 & 0.3676 & 0.3326 \\
			~ & video201 & 0.2763 & 0.3401 & 0.3042 \\
			~ & video210 & 0.2507 & 0.3918 & 0.3272 \\
			\hline
			\multirow{4}*{24} & video1 & 0.2680 & 0.3223 & 0.3039 \\
			~ & video40 & 0.2475 & 0.3236 & 0.2888 \\
			~ & video45 & 0.2120 & 0.2820 & 0.2611 \\
			~ & video82 & 0.2250 & 0.2702 & 0.2467 \\
			\hline
			\multirow{4}*{28} & video76 & 0.2652 & 0.3529 & 0.3085 \\
			~ & video213 & 0.2498 & 0.2857 & 0.2745 \\
			~ & video293 & 0.2169 & 0.3044 & 0.2773 \\
			~ & video334 & 0.2626 & 0.3250 & 0.3034 \\
			\hline
			\multirow{4}*{32} & video5 & 0.2732 & 0.2976 & 0.2852 \\
			~ & video94 & 0.1967 & 0.2466 & 0.2331 \\
			~ & video221 & 0.2655 & 0.2984 & 0.2818 \\
			~ & video324 & 0.2500 & 0.2918 & 0.2668 \\
			\hline
		\end{tabular}
	}
\end{table}  

\subsection{Limitation of Mean Pooling in Frame-wise Feature Aggregation}
\label{sec: limitation}
To demonstrate the weakness of mean pooling implementation in frame-wise feature aggregation, we conduct a comprehensive analysis of semantic similarities among pairwise caption-frame data on the MSR-VTT dataset. More specifically, we sample frames with the sampling rate of one frame per second (FPS) per video, followed by the semantic similarities calculation among single caption and multiple extracted frames. Here, we report the semantic similarities results at different frame lengths ranging from 12 to 32 in Table \ref{sim_comparison}, where the interval of frame lengths is set to 4, and a specific frame length contains 4 video clips. By analyzing the table, we note that regardless of the frame lengths for a single video, the semantic similarities among caption-frame pairs data fluctuate between the minimal and maximal similarity values. Besides, when the mean pooling method is adopted to generate video representation for similarity calculation, the obtained similarity is always inferior to the maximal value, resulting from the misleading information brought by semantic-irrelevant frame features. In the following subsections, we present techniques to mitigate the adverse effect of mean pooling in the improved similarity calculator.  

\subsection{Improved Similarity Calculator}
\label{sec: similarity calculator}
\begin{figure*}[!t]
	\centerline{\includegraphics[width=\textwidth]{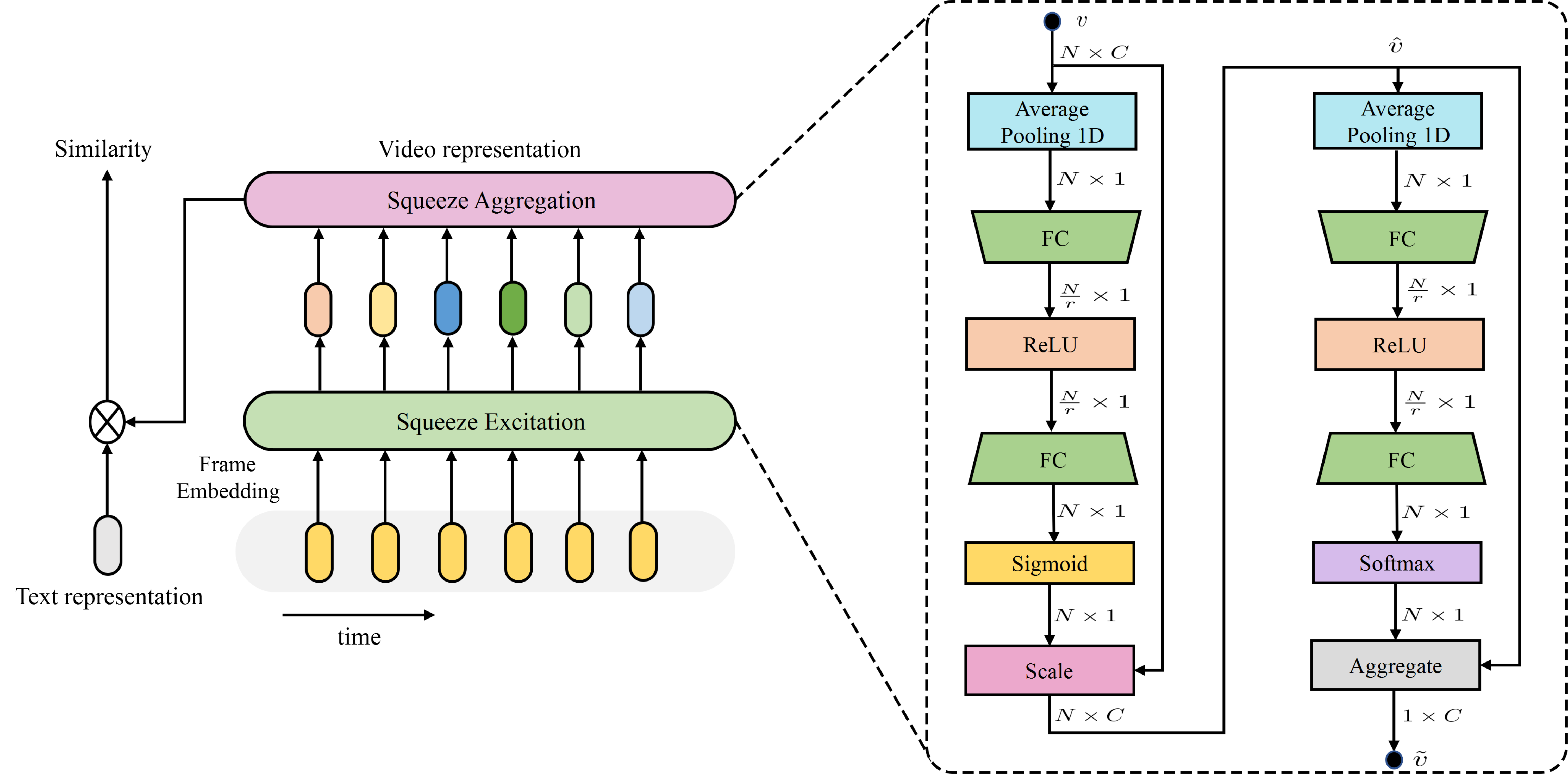}}
	\caption{Overview of the squeeze excitation-and-aggregation module. The extracted frame features are fed into squeeze excitation module to implement attentive frames enhancement and inattentive frames suppression, followed by the squeeze aggregation module to obtain the video representation. }
	\label{fig: architecture of squeeze ecitation-and-aggregation}
\end{figure*}

\begin{figure*}[!t]
	\centerline{\includegraphics[width=\textwidth]{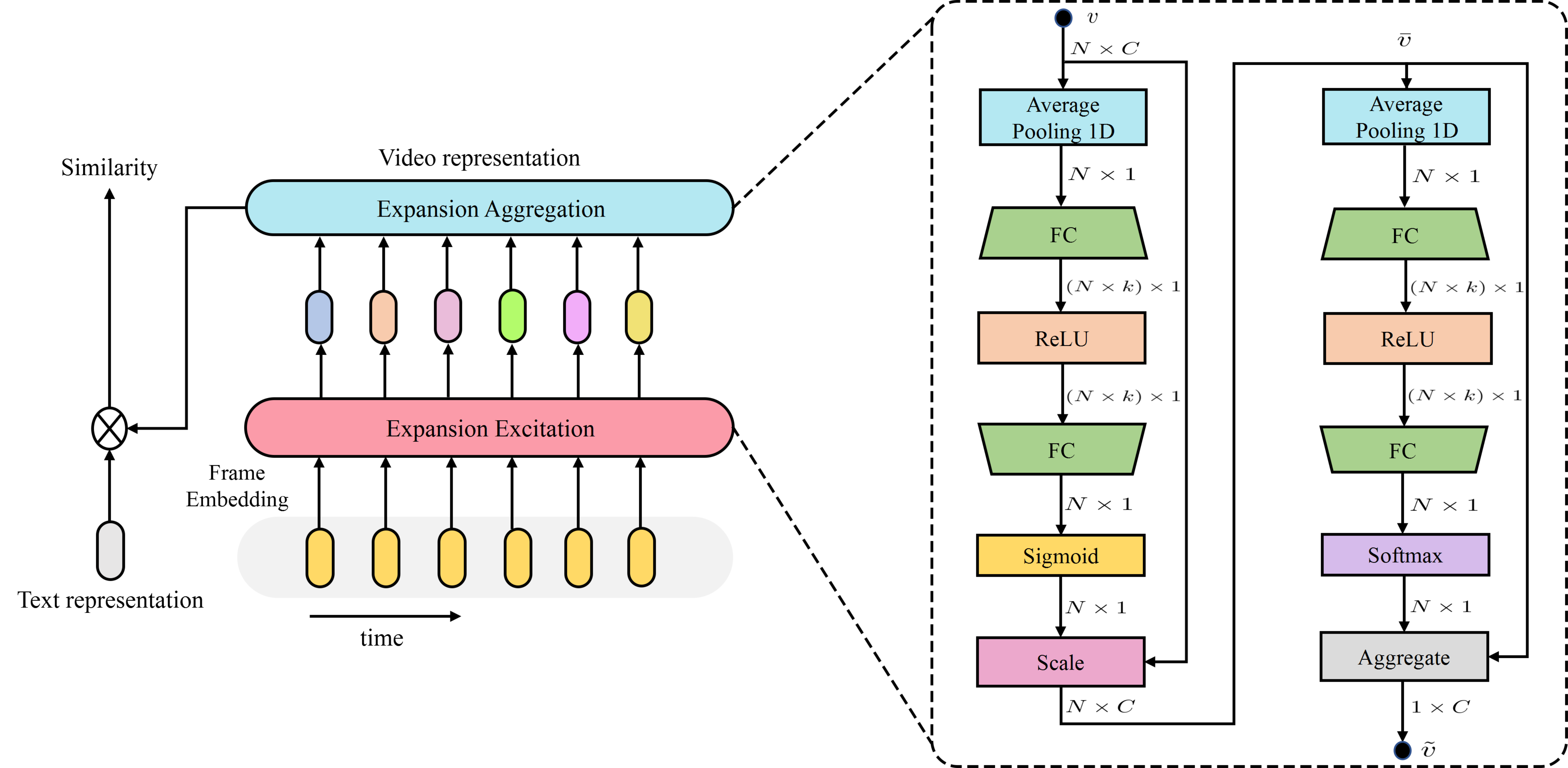}}
	\caption{Overview of the expansion excitation-and-aggregation module. The extracted frame features are fed into expansion excitation module to implement attentive frames enhancement and inattentive frames suppression, followed by the expansion aggregation module to obtain the video representation. }
	\label{fig: architecture of expansion ecitation-and-aggregation}
\end{figure*}

\begin{figure*}[!t]
	\centerline{\includegraphics[width=\textwidth]{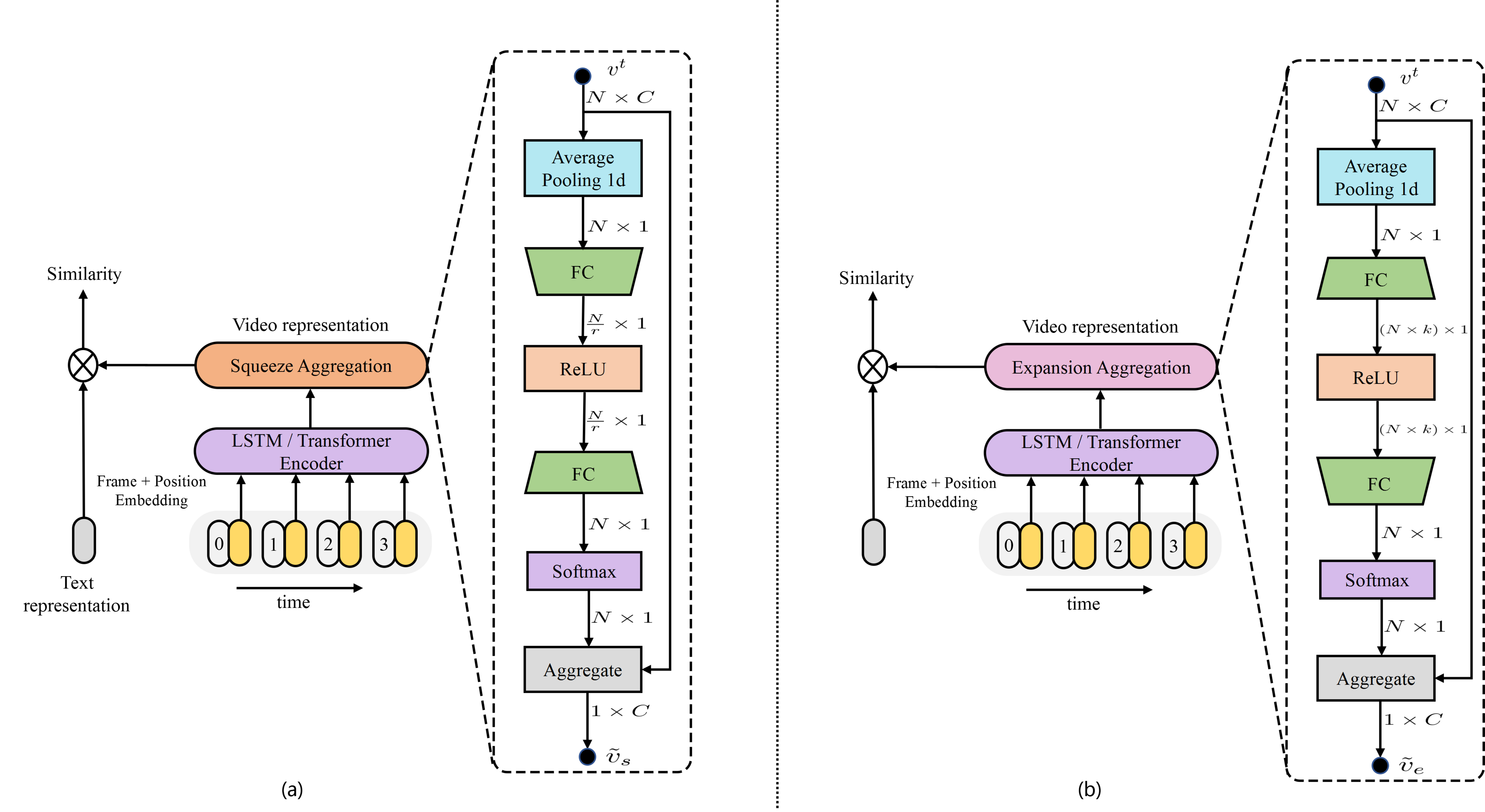}}
	\caption{(a) The diagram of squeeze aggregation module. (b) The diagram of expansion aggregation module. Both aggregation modules obtain the frame-wise weights through two fully-connected layers and a nonlinear activation function in between, followed by a weighted summation operation to aggregate the frame features into video representation. }
	\label{fig: architecture of squeeze and expansion aggregation}
\end{figure*}

\begin{figure*}[!t]
	\centerline{\includegraphics[width=\textwidth]{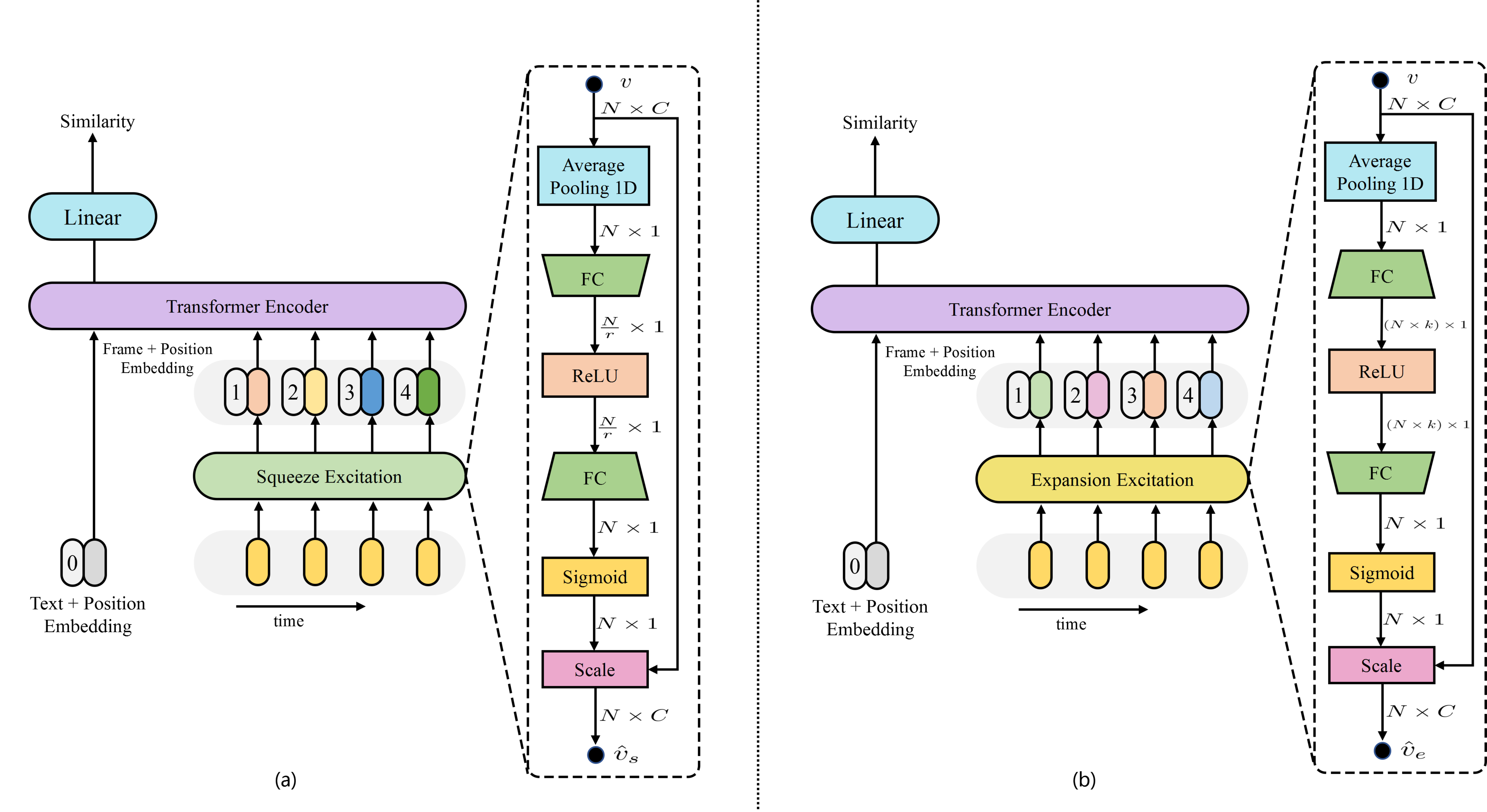}}
	\caption{(a) The overall architecture of squeeze excitation module. (b) The overall architecture of expansion excitation module. By explicitly modeling relationships among distinct frames, it achieves frame-wise feature recalibration. Then, the textual and visual representations are fed into Transformer Encoder for multi-modal interaction to compute the similarity score. }
	\label{fig: architecture of squeeze_and_expansion excitation}
\end{figure*}

\begin{table}
	\caption{Network settings of excitation-and-aggregation modules in squeeze and expansion paradigms on the MSR-VTT dataset. Input frame features size is $12 \times 512$ by default. The squeeze ratio $r$ and expansion ratio $k$ are set to 4. }
	\label{table: Excitation-and-Aggregation Network Settings}
	\centering
	\resizebox{0.5\textwidth}{!}{
	\begin{tabular}{c|c|c|c}
		\hline
	     \multirow{2}*{Stage} & \multirow{2}*{Layer Name} & \multicolumn{2}{c}{Output Size} \\
	     \cline{3-4}
	     & & Squeeze & Expansion \\
		\hline
		\multirow{6}*{Excitation} & Average Pooling (1D) & $12 \times 1$ & $12 \times 1$\\
		~ & FC & $3 \times 1$ $\downarrow$ & $48 \times 1$ $\uparrow$\\
		~ & ReLU & $3 \times 1$ & $48 \times 1$ \\
		~ & FC & $12 \times 1$ $\uparrow$ & $12 \times 1$ $\downarrow$ \\
		~ & Sigmoid & $12 \times 1$ & $12 \times 1$ \\
		~ & Scale & $12 \times 512$ & $12 \times 512$ \\
		\hline
		\multirow{6}*{Aggregation} & Average Pooling (1D) & $12 \times 1$ & $12 \times 1$ \\
		~ & FC & $3 \times 1$ $\downarrow$ & $48 \times 1$ $\uparrow$ \\
		~ & ReLU & $3 \times 1$ & $48 \times 1$ \\
		~ & FC & $12 \times 1$ $\uparrow$ & $12 \times 1$ $\downarrow$ \\
		~ & Softmax & $12 \times 1$ & $12 \times 1$ \\
		~ & Aggregate & $1 \times 512$ &  $1 \times 512$\\
		\hline
	\end{tabular}
	}
\end{table}

\begin{table}
	\caption{Network settings of aggregation and excitation modules in squeeze and expansion paradigms on the MSR-VTT dataset. Input frame features size is $12 \times 512$ by default. The squeeze ratio $r$ and expansion ratio $k$ are set to 4. }
	\label{table: Aggregation Network Settings}
	\centering
	\resizebox{0.5\textwidth}{!}{
		\begin{tabular}{c|c|c|c}
			\hline
			\multirow{2}*{Module} & \multirow{2}*{Layer Name} & \multicolumn{2}{c}{Output Size} \\
			\cline{3-4}
			&& Squeeze & Expansion \\
			\hline
			\multirow{6}*{\makecell[c]{Aggregation\\(Sequential Type)}} & Average Pooling (1D) & $12 \times 1$ & $12 \times 1$ \\
			~ & FC & $3 \times 1$ $\downarrow$ & $48 \times 1$ $\uparrow$ \\
			~ & ReLU & $3 \times 1$ & $48 \times 1$ \\
			~ & FC & $12 \times 1$ $\uparrow$ & $12 \times 1$ $\downarrow$ \\
			~ & Softmax & $12 \times 1$ & $12 \times 1$ \\
			~ & Aggregate & $1 \times 512$ &  $1 \times 512$ \\
			\hline
			\multirow{6}*{\makecell[c]{Excitation\\(Tight Type)}} & Average Pooling (1D) & $12 \times 1$ & $12 \times 1$ \\
			~ & FC & $3 \times 1$ $\downarrow$ & $48 \times 1$ $\uparrow$ \\
			~ & ReLU & $3 \times 1$ & $48 \times 1$ \\
			~ & FC & $12 \times 1$ $\uparrow$ & $12 \times 1$ $\downarrow$ \\
			~ & Sigmoid & $12 \times 1$ & $12 \times 1$ \\
			~ & Scale & $12 \times 512$ &  $12 \times 512$ \\
			\hline
		\end{tabular}
	}
\end{table}

As depicted in Fig. \ref{fig: main_structure}, the architecture consists of three critical components, termed Video Encoder, Text Encoder and Improved Similarity Calculator, of which our main efforts are focused on the structure improvement of similarity calculator. The following subsections give detail descriptions of excitation-and-aggregation, aggregation and excitation modules in the squeeze and expansion paradigms respectively.

\subsubsection{Excitation-and-Aggregation Module} Instead of simply averaging frame features into video representation, we propose the excitation-and-aggregation modules, including squeeze and expansion paradigms. The details of network settings are reported in Table \ref{table: Excitation-and-Aggregation Network Settings}.

\textbf{Squeeze Paradigm. } As shown in the Fig. \ref{fig: architecture of squeeze ecitation-and-aggregation}, the module contains squeeze excitation and squeeze aggregation module, of which the former is used for frame-wise features recalibration and the latter is adapted to frame features aggregation.

To handle 1D sequence input, the module firstly utilizes 1D average pooling to squeeze extracted frame features $v\in \mathbb{R}^{N \times C}$ along channel dimension and generate frame-wise statistics $u\in \mathbb{R}^{N \times 1}$, where $N$ is the length of video frames, and $C$ is the channel of frame feature. The $i$-th element of $u$ is computed through:
\begin{equation}
	u_{i}=\frac{1}{C}\sum_{j=1}^{C}v_{ij}, \forall i \in [1, N].
\end{equation} 
Then a simple gating mechanism with a sigmoid activation is implemented to map the aforementioned frame statistics into a set of frame weights. Specifically, the gating mechanism is formed into a bottleneck layer, consisting of two fully-connected (FC) layers around the ReLU activation function. The first layer reduces dimension with reduction ratio $r$, followed by a nonlinear activation, and last layer returns to the dimension of average pooling output $u$. And the activation $z \in \mathbb{R}^{N \times 1}$ is calculated by sigmoid activation:
\begin{equation}
	z=\delta(W_{2}\sigma(W_{1}u+b_{1})+b_{2}),
\end{equation}
where $W_{1} \in \mathbb{R}^{\frac{N}{r} \times N}$ and $W_{2} \in \mathbb{R}^{ N \times \frac{N}{r}}$ refer to the weights of two FC layers. $b_{1} \in \mathbb{R}^{\frac{N}{r}}$ and $b_{2} \in \mathbb{R}^{N}$ are the biases. And $\sigma(\cdot)$ and $\delta(\cdot)$ denote the ReLU function and sigmoid activation. \\
Finally, the input $v \in \mathbb{R}^{N \times C}$ is scaled with obtained activations:
\begin{equation}
	\hat{v}_{ij}=z_{i}v_{ij}, \forall i \in [1, N], \forall j \in [1, C],
\end{equation}
where the frame-wise multiplication between scalar $z_{i}$ and frame feature $v_{i}$ is used to scale frame representations.

Unlike sigmoid function and scaling operation in the squeeze excitation module, squeeze aggregation module is built by the utilization of softmax function and aggregation operation. As a complementation to sigmoid function in the excitation module, the softmax function is capable of modeling mutually exclusive relationships among frame features, and the generated frame-wise weights are used for frame-level features aggregation. The procedures can be noted as:
\begin{align}
	\hat{u}_{i}&=\frac{1}{C}\sum_{j=1}^{C}\hat{v}_{ij}, \forall i \in [1, N],  \label{4} \\
	\hat{z}&=s(W_{4}\sigma(W_{3}\hat{u}+b_{3})+b_{4}) \label{5}, \\ 
	\tilde{v}&=\sum_{i=1}^{N} \hat{z}_{i} \hat{v}_{i} \label{6},
\end{align}
where $\hat{u}_{i}$ is the $i$-th element of output from 1D average pooling in Eq. \eqref{4}. In Eq. \eqref{5}, $s$ is the softmax function, and $\{W_{3}, b_{3}, W_{4}, b_{4}\}$ are learnable parameters in two FC layers. $\tilde{v}$ in Eq. \eqref{6} refers to the obtained video representation. 

\textbf{Expansion paradigm. } The module architecture is illustrated in Fig. \ref{fig: architecture of expansion ecitation-and-aggregation}. For given frame features $v\in \mathbb{R}^{N \times C}$, we obtain video representation $\tilde{v} \in \mathbb{R}^{N \times C}$ through following formulations:
\begin{align}
	\bar{u}_{i}&=\frac{1}{C}\sum_{j=1}^{C}v_{ij}, \forall i \in [1, N], \\
	\bar{z}&=\delta(W_{6}\sigma(W_{5}\bar{u}+b_{5})+b_{6}), \\
	\bar{v}_{ij}&=\bar{z}_{i}v_{ij}, \forall i \in [1, N], \forall j \in [1, C], \\
	\tilde{u}_{i}&=\frac{1}{C}\sum_{j=1}^{C}\bar{v}_{ij}, \forall i \in [1, N], \\
	\tilde{z}&=s(W_{8}\sigma(W_{7}\tilde{u}+b_{7})+b_{8}), \\
	\tilde{v}&=\sum_{i=1}^{N} \tilde{z}_{i} \bar{v}_{i},
\end{align}
where $W_{7} \in \mathbb{R}^{(N \times k) \times N}$ and $W_{8} \in \mathbb{R}^{ N \times (N \times k)}$ refer to the weights of two FC layers. $b_{7} \in \mathbb{R}^{N \times k}$ and $b_{8} \in \mathbb{R}^{N}$ are the corresponding biases. $k$ denotes the expansion ratio.

\subsubsection{Aggregation Modules in the sequential type} Due to the lack of temporal information consideration in the pure mean pooling method, the CLIP4Clip introduces sequential LSTM and Transformer Encoder to achieve relationship modeling among frame representations. To further improve the performance of sequential modules, we propose aggregation modules shown in Fig. \ref{fig: architecture of squeeze and expansion aggregation}, including squeeze and expansion paradigms, to aggregate temporal frame features into video feature. The details of networks settings with different paradigms are given in the top part of Table \ref{table: Aggregation Network Settings}.

\textbf{Squeeze Paradigm. } Given frame features $v^{t} \in \mathbb{R}^{N \times C}$ output from sequential modules, we firstly average features across frame dimension through 1D average pooling into frame-wise statistics:
\begin{equation}
	(u_{s}^{t})_{i}=\frac{1}{C}\sum_{j=1}^{C} v^{t}_{ij}, \forall i \in [1, N],
\end{equation}
where $(u_{s}^{t})_{i}$ is the $i$-th element of $u_{s}^{t}$. \\
Then two FC layers are adapted to dimension reduction and expansion, and the ReLU activation function is utilized in between. Through the softmax function, frame-wise weights $z_{s}^{t} \in \mathbb{R}^{N \times 1}$ are computed as:
\begin{equation}
	z_{s}^{t}=s(W_{10}\sigma(W_{9}u_{s}^{t}+b_{9})+b_{10}),
\end{equation}
where $\{W_{9} \in \mathbb{R}^{\frac{N}{r} \times N}, b_{9} \in \mathbb{R}^{\frac{N}{r}}\}$ are weight and bias in the first FC layer, and $\{W_{10} \in \mathbb{R}^{N \times \frac{N}{r}}, b_{10} \in \mathbb{R}^{N}\}$ are same type parameters of the second FC layer, of which $r$ is the reduction ratio. The final video representation is derived by a weighted sum over input frame features:
\begin{equation}
	\tilde{v}_{s}=\sum_{i=1}^{N} (z_{s}^{t})_{i}  v^{t}_{i}. 
\end{equation}

\textbf{Expansion Paradigm. } Similar to the squeeze aggregation module, we obtain the video representation through following formulations:
\begin{align}
	(u_{e}^{t})_{i}&=\frac{1}{C}\sum_{j=1}^{C} v^{t}_{ij}, \forall i \in [1, N], \\
	z_{e}^{t}&=s(W_{12}\sigma(W_{11}u_{s}^{t}+b_{11})+b_{12}), \\
	\tilde{v}_{e}&=\sum_{i=1}^{N} (z_{e}^{t})_{i}  v^{t}_{i},
\end{align}
where $W_{11} \in \mathbb{R}^{(N \times k) \times N}$ and $W_{12} \in \mathbb{R}^{ N \times (N \times k)}$ denote the weights of two FC layers. $b_{11} \in \mathbb{R}^{N \times k}$ and $b_{12} \in \mathbb{R}^{N}$ are the biases. $k$ refers to the expansion ratio.

\subsubsection{Excitation Modules in the tight type} We propose excitation modules to achieve relationships modeling among frame features, followed by the multi-modal interaction between textual and visual representations. Fig. \ref{fig: architecture of squeeze_and_expansion excitation} shows the architecture of squeeze and expansion excitation modules in the tight type. We report the details of network settings in the bottom part of Table \ref{table: Aggregation Network Settings}.

\textbf{Squeeze Paradigm. } For extracted frame features $v \in \mathbb{R}^{N \times C}$, the excitation frame representations $\hat{v}_{s} \in \mathbb{R}^{N \times C}$ are computed as:
\begin{align}
	(\bar{u}_{s})_{i}&=\frac{1}{C}\sum_{j=1}^{C}v_{ij}, \forall i \in [1, N], \label{10} \\
	\bar{z}_{s}&=\delta(W_{14}\sigma(W_{13}\bar{u}+b_{13})+b_{14}) \label{11}, \\ 
	(\hat{v}_{s})_{ij}&= (\bar{z}_{s})_{i} v_{ij}, \forall i \in [1, N], \forall j \in [1, C], \label{12}
\end{align}
where $(\bar{u}_{s})_{i}$ is the $i$-th element of output from 1D average pooling in Eq. \eqref{10}. The parameters $\{W_{13} \in \mathbb{R}^{\frac{N}{r} \times N}, b_{13} \in \mathbb{R}^{\frac{N}{r}}\}$ and $\{W_{14} \in \mathbb{R}^{N \times \frac{N}{r}}, b_{14} \in \mathbb{R}^{N}\}$ in Eq. \eqref{11} refer to weights and biases in the two FC layers. In Eq. \eqref{12}, $\hat{v}_{s} \in \mathbb{R}^{N \times C}$ denotes the discriminative frame features used for multi-modal Transformer Encoder.

\textbf{Expansion Paradigm. } Since the architecture of expansion paradigm is similar to the squeeze paradigm, we directly give the following formulations to obtain the excitation frame representations $\hat{v}_{e} \in \mathbb{R}^{N \times C}$:
\begin{align}
	(\bar{u}_{e})_{i}&=\frac{1}{C}\sum_{j=1}^{C}v_{ij}, \forall i \in [1, N], \\
	\bar{z}_{e}&=\delta(W_{16}\sigma(W_{15}\bar{u}+b_{15})+b_{16}) , \\ 
	(\hat{v}_{e})_{ij}&= (\bar{z}_{e})_{i} v_{ij}, \forall i \in [1, N], \forall j \in [1, C],  
\end{align}
where $\{W_{15} \in \mathbb{R}^{(N \times k) \times N}, W_{16} \in \mathbb{R}^{ N \times (N \times k)}\}$ refer to the weights of two FC layers, $\{b_{15} \in \mathbb{R}^{N \times k}, b_{16} \in \mathbb{R}^{N}\}$ are the corresponding biases. $k$ denotes the expansion ratio.

\section{Experiment}
\label{Experiment}
\subsection{Datasets}
\textbf{MSR-VTT} \cite{xu2016msr} is a popular dataset containing 10k video clips, each with a length between 10 and 32 seconds. For a single video clip, 20 different captions are labeled manually and 20 clip-caption pairs are generated. In this paper, we adopt the widely-used ``Training-9K'' data splits in our experiments if no extra annotation, where the training and testing data contain 9k and 1k videos respectively. \par
\textbf{ActivityNet} \cite{caba2015activitynet} contains 20k YouTube videos and 100k caption annotations. We concatenate all the descriptions of a given video to form one paragraph, which is used for model evaluation in the video-paragraph retrieval paradigm. \par
\textbf{DiDeMo} \cite{anne2017localizing} consists of 10k Flickr videos annotated with 40k captions. Following previous works \cite{luo2022clip4clip, gabeur2020multi, zhang2022local}, all captions of a video are concatenated to generate the single paragraph query during video-paragraph retrieval evaluation.  

\subsection{Evaluation Metrics}
To evaluate the retrieval performance of our proposed modules, we use the standard evaluation metrics, including recall at Rank K (R@K, higher is better), median rank (MdR, lower is better) and mean rank (MnR, lower is better). 

R@K is defined as the percentage of correct matches in the top-K retrieved results to the query instance:
\begin{equation}
	\text{R}@\text{K}=100 \times \frac{n}{N_{k}},
\end{equation}
where $n$ is the number of correct matches and $N_{k}$ is the number of top-K retrieved results.

MdR and MnR calculate the median and average rank of all correct results respectively, which can be represented as follows:
\begin{align}
	\text{MdR}&= \text{Median}(rank_{1}, \cdots, rank_{i}, \cdots, rank_{n}), \label{14} \\
	\text{MnR}&= \text{Mean}(rank_{1}, \cdots, rank_{i}, \cdots, rank_{n}), \label{15} 
\end{align}
where $rank_{i}$ refers to the rank of $i$-th element in correct results.

To measure the overall retrieval performance, we also sum the all R@K results as RSum, which can be computed as:
\begin{equation}
	\text{RSum}=\sum\nolimits_{\text{K }\in \text{A}} \text{R@K},
\end{equation}
where $\text{A}=\{1,5,10\}$ is the set of K values.

\subsection{Implementation Details}
In this paper, we reuse parameters from the pre-trained CLIP (ViT-B/32) for better parameters initialization. Specifically, in the sequential and tight types, the position embedding from CLIP's text encoder is repeated to initialize the position embedding in transformer encoder. The weights in CLIP's image encoder are utilized for parameter initialization in the vision transformer encoder. The parameters in linear projection layer are randomly initialized. The layer of sequential modules, including LSTM and Transformer Encoder are set to 1 and 4 in our experiments. For simplicity, we use 2D patch linear, of which the resolution of each frame patch is set to 32, to flatten frame patches into 1D frame sequence.  

To finetune the aforementioned datasets, a cosine schedule is applied to decay the learning rate, and the Adam optimizer is used in the model optimization. For the CLIP pre-trained text encoder and image encoder, the initial learning rate is set to 1e-7. Whereas in the new modules, e.g. Transformer Encoder and aforementioned modules, we set a larger initial learning rate of 1e-4. During MSR-VTT experiments, we set the training epoch, batch size, frame length and caption token length to 5, 128, 12, and 32. Due to the longer and more complex videos and captions in the ActivityNet and DiDeMo, the training epoch, frame length and caption length are promoted to 20, 64, and 64 for comprehensive video-paragraph reasoning. Limited to the GPU memory, the batch size of ActivityNet and DiDeMo are reduced to 64. All experiments are carried out on 4 NVIDIA GeForce RTX 3090 24GB GPUs based on Pytorch library. In order to speed up the training process, videos are compressed to 3 frames-per-second (FPS) with the resolution of 224 in height or width.

\subsection{Performance Comparison}
To validate the effectiveness of our proposed modules, we compare the retrieval performance with previous methods on the MSR-VTT, ActivityNet and DiDeMo datasets. The utilization of these modules achieves better retrieval results on three datasets with substantial improvements. It is worth noting that due to the incomplete results presentation on specific datasets in some approaches, several experimental data are not reported in the results, including inference speed and certain evaluation metrics like MdR and MnR. \par

\subsubsection{MSR-VTT Results} 
\noindent\textbf{Retrieval Performance.} By analyzing the retrieval results in Table \ref{table: MSR-VTT Performance}, we observe that
compared with the rough mean pooling implementation, our proposed modules bring significant performance gain in  three types. For the pure mean pooling type, the squeeze excitation-and-aggregation module achieves 43.5 R@1 and 195.6 RSum, bringing +0.9\% (+0.4\%) and +0.6\% (+1.1\%) relative (absolute) performance improvements. For the sequential type, the combination of expansion aggregation module and sequential modules, include LSTM and Transformer Encoder, outperforms T2V R@1 in CLIP4Clip by +2.6\% and +1.2\% relative improvements. Besides, when the squeeze excitation module is adopted in the tight type, the better R@1 of 41.5 and RSum of 195.4 are obtained, demonstrating the superiority of squeeze excitation module in the multi-modal interaction enhancement. 

\noindent\textbf{Computational Overhead.} For the inference time in Table \ref{table: MSR-VTT Performance}, we show that the proposed excitation-and-aggregation designs increase computation costs among three types. More precisely, when the squeeze excitation-and-aggregation module is used in frame features aggregation, the inference time is closely doubled. For the sequential type, the integration of sequential module and expansion aggregation module extends the inference time by +15.6\% and +0.5\% respectively. For the tight type, it is noticeable that the multi-modal interaction takes 582.0 milliseconds during per video evaluation, and the inference time is increased to 590.3 milliseconds with the introduction of squeeze excitation module. Though the proposed method needs more inference time, the significant improvement of overall retrieval performance with slight computation cost increase is acceptable.      

\subsubsection{ActivityNet Results} 
\noindent\textbf{Retrieval Performance.} The ActivityNet results are given in Table \ref{table: ActivityNet Performance}. It can be seen that our proposed expansion excitation-and-aggregation module achieves T2V R@1 of 42.4 and RSum of 202.0, outperforming the results in CLIP4Clip-meanP by +3.2\% (+1.3\%) and +1.6\% (+3.2\%) relative (absolute) performance improvements. We also observe the strength of expansion aggregation module, which obtains T2V R@1 of 42.6 and 44.1 with +3.9\% (+1.6\%) and +4.5\% (+1.9\%) relative (absolute) improvements in comparison to the pure sequential LSTM and Transformer Encoder. Moreover, it is noticeable that the adoption of expansion excitation module in the tight type is a better alternative for the video-paragraph retrieval task, which boosts the T2V R@1 and RSum by +5.4\% and +2.7\%. This significant performance gain can be attributed to two aspects. On the one hand, our proposed excitation and aggregation modules are critical to reducing the bad effects of inattentive frames and enhancing the visual information related to the paragraph query. On the other hand, benefiting from the nonlinear enhancement introduced by the sigmoid activation function, the model can better adapt to complex relationships and nonlinear features in the video-paragraph retrieval task.

\noindent\textbf{Computational Overhead.} Table \ref{table: ActivityNet Performance} reports the inference time of excitation and aggregation designs during video-paragraph retrieval evaluation in the three types. We observe that although the inference time is almost doubled compared with the MSR-VTT results, our proposed modules increase slight inference time on the order of tens of milliseconds with significant retrieval performance improvement. For example, when the expansion aggregation module is used followed by the Transformer Encoder, the negligible extra inference time of 3.5 milliseconds boosts the T2V performance of R@1 and RSum by +1.9\% and +3.9\% absolute improvements, which demonstrates the added computational overhead is deserving. 

\subsubsection{DiDeMo Results} 
\noindent\textbf{Retrieval Performance.} As shown in Table \ref{table: DiDeMo Performance}, we can see that the proposed expansion excitation-and-aggregation module achieves lower T2V 40.1 R@1 and 183.9 RSum with 0.7\% and 2.1\% absolute performance decline. Compared with the pure sequential LSTM and Transformer Encoder, the retrieval performance also suffers a great decrease. For example, the cascade of expansion aggregation module and sequential LSTM achieves lower T2V R@1 of 38.8, leading to 2.0\% absolute decline. The reason behind this abnormal decrease is thoroughly elaborated in the Appendix Section \ref{app: performance comparison}. Different from the significant performance decline brought by the former three designs, we observe that the introduction of expansion excitation module in the tight type achieves better 31.0 R@1 and 162.5 RSum in the text-video retrieval task, outperforming the baseline results by +9.5\% (+2.7\%) and +3.5\% (+5.5\%) relative (absolute) improvements.

\noindent\textbf{Computational Overhead.} Due to the sole performance improvement given by the expansion excitation module, we mainly analyze the trade-off between performance gain and increased inference time in the tight type. Through the negligible increase inference cost of 18.4 milliseconds, the expansion excitation module promotes the T2V and V2T R@1 to 31.0 and 34.6 respectively with +2.7\% and +2.1\% absolute improvements, verifying the additional computational overhead is valuable.

\begin{table*}
	\caption{Retrieval performance comparison on the MSR-VTT dataset. Speed is the inference time per video during evaluation on a NVIDIA GeForce RTX 3090 GPU. $\dag$ indicates that results are obtained by our re-training. } 
	\label{table: MSR-VTT Performance}
	\centering
	\resizebox{\textwidth}{!}{
		\renewcommand{\arraystretch}{1.05} 
		\begin{tabular}{l|c|cccccc|cccccc} 
			\hline 
			\multirow{2}*{Methods}  & \multirow{2}*{\makecell[c]{Speed\\(ms)}} & \multicolumn{6}{c|}{Text-to-Video} &  \multicolumn{6}{c}{Video-to-Text} \\ 
			\cline{3-8} \cline{9-14}  
			&& R@1$\uparrow$  & R@5$\uparrow$  & R@10$\uparrow$ & MdR$\downarrow$ &  MnR$\downarrow$ &
			RSum$\uparrow$
			& R@1$\uparrow$  & R@5$\uparrow$  & R@10$\uparrow$ & MdR$\downarrow$ &  MnR$\downarrow$ & 
			RSum$\uparrow$ \\
			\hline 
			$\textup{CE}$ \cite{liu2019use} & - & 20.9 & 48.8 & 62.4 & 6.0 & 28.2 & 132.1 & 20.6 & 50.3 & 64.0 & 5.3 & - & 134.9 \\ 
			$\textup{MMT}$ \cite{gabeur2020multi} & - & 26.6 & 57.1 & 69.6 & 4.0 & 24.0 & 153.3 & 27.0 & 57.5 & 69.7 & 3.7 & - & 154.2 \\
			$\textup{SSB}$ \cite{patrick2020support} & - & 30.1 & 58.5 & 69.3 & 3.0 & - & 157.9 & 28.5 & 58.6 & 71.6 & 3.0 & - & 158.7 \\
			$\textup{MDMMT}$ \cite{dzabraev2021mdmmt} & - & 38.9 & 69.0 & 79.7 & 2.0 & 16.5 & 187.6 & - & - & - & - & - & - \\
			$\textup{HiT}$ \cite{liu2021hit}  & - & 30.7 & 60.9 & 73.2 & 2.6 & - & 164.8 & 32.1 & 62.7 & 74.1 & 3.0 & - & 168.9 \\
			$\textup{TT-CE+}$ \cite{croitoru2021teachtext}  & - & 29.6 & 61.6 & 74.2 & 3.0 & - & 165.4 & 32.1 & 62.7 & 75.0 & 3.0 & - & 169.8 \\
			\hline
			$\textup{(CLIP4Clip)-meanP} \dag$ \cite{luo2022clip4clip} & 14.3 & 43.1 & 70.8 & 80.6 & \textbf{2.0} & 17.0 & 194.5 & \textbf{42.6} & 70.0 & 80.6 & \textbf{2.0} & 12.1 & 193.2 \\
			$\textup{(Ours)-Squeeze Excitation-and-Aggregation}$ &  27.1 & \textbf{43.5} & \textbf{71.1} & \textbf{81.0} & \textbf{2.0} & \textbf{15.7} & \textbf{195.6} & 42.3 & \textbf{70.8} & \textbf{80.9} & \textbf{2.0} & \textbf{11.7} & \textbf{194.0} \\
			\hline
			$\textup{(CLIP4Clip)-seqLSTM} \dag$ \cite{luo2022clip4clip} & 52.4 & 42.4 & 69.4 & 80.1 & \textbf{2.0} & 17.2 & 191.9 & 40.9 & \textbf{69.8} & 79.5 &  \textbf{2.0} & 13.0 & 190.2 \\
			$\textup{(Ours)-seqLSTM + Expansion Aggregation}$ & 60.6 & \textbf{43.5} & \textbf{70.5} & \textbf{80.8} & \textbf{2.0} & \textbf{16.4} & \textbf{194.8} & \textbf{42.2} & 69.7 & \textbf{81.2} & \textbf{2.0} & \textbf{11.9} & \textbf{193.1} \\
			\hline
			$\textup{(CLIP4Clip)-seqTransf} \dag$ \cite{luo2022clip4clip} & 199.3 & 43.4 & \textbf{71.8} & 81.3 & \textbf{2.0} & 16.4 & 196.5 & \textbf{44.1} & \textbf{70.6} & 80.0 &  \textbf{2.0} & 12.0 & \textbf{194.7} \\
			$\textup{(Ours)-seqTransf + Expansion Aggregation}$ & 204.3 & \textbf{43.9} & 71.2 & \textbf{82.0} & \textbf{2.0} & \textbf{15.9} & \textbf{197.1} & 43.7 & 69.9 & \textbf{80.3} & \textbf{2.0} & \textbf{11.8} & 193.9 \\ 
			\hline
			$\textup{(CLIP4Clip)-tightTransf} \dag$ \cite{luo2022clip4clip} & 582.0 & 40.4 & \textbf{72.3} & 82.0 & \textbf{2.0} & \textbf{13.3} & 194.7 & 40.7 & 70.1 & \textbf{80.3} & \textbf{2.0} & \textbf{12.3} & \textbf{191.1} \\
			$\textup{(Ours)-tightTransf + Squeeze Excitation}$ & 590.3 & \textbf{41.5} & 71.6 & \textbf{82.3} & \textbf{2.0} & 13.5 & \textbf{195.4} & \textbf{41.2} & \textbf{70.3} & 79.4 & \textbf{2.0} & 13.1 & 190.9 \\
			\hline
		\end{tabular}
	}
\end{table*}

\begin{table*}
	\caption{Retrieval performance comparison on the ActivityNet dataset. Speed is the inference time per video during evaluation on a NVIDIA GeForce RTX 3090 GPU. $\dag$ indicates re-training.}
	\label{table: ActivityNet Performance}
	\centering
	\resizebox{\textwidth}{!}{
		\renewcommand{\arraystretch}{1.05} 
		\begin{tabular}{l|c|cccccc|cccccc}
			\hline \rule{0pt}{8pt}
			\multirow{2}*{Methods} & \multirow{2}*{\makecell[c]{Speed\\(ms)}} & \multicolumn{6}{c|}{Text-to-Video} &  \multicolumn{6}{c}{Video-to-Text} \\ 
			\cline{3-8} \cline{9-14}  
			&& R@1$\uparrow$  & R@5$\uparrow$  & R@10$\uparrow$ & MdR$\downarrow$ &  MnR$\downarrow$ & 
			RSum$\uparrow$
			& R@1$\uparrow$  & R@5$\uparrow$  & R@10$\uparrow$ & MdR$\downarrow$ &  MnR$\downarrow$ & 
			RSum$\uparrow$ \\
			\hline
			$\textup{FSE}$ \cite{zhang2018cross} & - & 18.2 & 44.8 & - & 7.0 & - & - &  16.7 & 43.1 & - & 7.0 & - & - \\
			$\textup{CE}$ \cite{liu2019use} & - & 18.2 & 47.7 & - & 6.0 & 23.1 & - & 17.7 & 46.6 & - & 6.0 & 24.4 & - \\ 
			$\textup{MMT}$ \cite{gabeur2020multi} & - & 28.7 & 61.4 & - & 3.3 & 16.0 & - & 28.9 & 61.1 & - & 4.0 & 17.1 & -  \\
			$\textup{SSB}$ \cite{patrick2020support} & - & 29.2 & 61.6 & - & 3.0 & - & - & 28.7 & 60.8 & - & 2.0 & - & - \\
			$\textup{ClipBERT}$ \cite{lei2021less} & - & 21.3 & 49.0 & - & 6.0 & - & - & - & - & - & - & - \\
			$\textup{TT-CE+}$ \cite{croitoru2021teachtext} & - & 23.5 & 57.2 & - & 4.0 & - & - & 23.0 & 56.1 & - & 4.0 & - & - \\
			\hline
			$\textup{(CLIP4Clip)-meanP} \dag$ \cite{luo2022clip4clip} & 37.7 & 41.1 & 73.1 & 84.6 & \textbf{2.0} & \textbf{7.6} & 198.8 & 42.1 & 73.7 & 85.4 & \textbf{2.0} & 7.1 & 201.2 \\
			$\textup{(Ours)-Expansion Excitation-and-Aggregation}$ & 75.0 & \textbf{42.4} & \textbf{74.3} & \textbf{85.3} & \textbf{2.0} & 7.8 & \textbf{202.0} & \textbf{44.2} & \textbf{75.3} & \textbf{87.0} & \textbf{2.0} & \textbf{6.5} & \textbf{206.5}  \\
			\hline
			$\textup{(CLIP4Clip)-seqLSTM} \dag$ \cite{luo2022clip4clip} & 292.5 & 41.0 & 73.0 & 84.2 & \textbf{2.0} & \textbf{7.7} & 198.2 & 41.9 & 73.0 & 85.2 & \textbf{2.0} & 7.1 & 200.1 \\
			$\textup{(Ours)-seqLSTM + Expansion Aggregation}$ & 315.2 & \textbf{42.6} & \textbf{74.5} & \textbf{85.3} & \textbf{2.0} & 8.0 & \textbf{202.4} & \textbf{44.6} & \textbf{75.1} & \textbf{86.6} & \textbf{2.0} & \textbf{6.7} & \textbf{206.3} \\
			\hline
			$\textup{(CLIP4Clip)-seqTransf} \dag$  \cite{luo2022clip4clip} & 959.2 & 42.2 & 74.1 & 85.2 & \textbf{2.0} & 7.2 & 201.5 & 41.9 & 73.5 & 85.6 & \textbf{2.0} & 7.2 & 201.0 \\
			$\textup{(Ours)-seqTransf + Expansion  Aggregation}$ & 962.7 & \textbf{44.1} & \textbf{75.1} & \textbf{86.2} & \textbf{2.0} & \textbf{7.1} & \textbf{205.4} & \textbf{43.9} & \textbf{75.0} & \textbf{86.6} & \textbf{2.0} & \textbf{6.6} & \textbf{205.5} \\
			\hline
			$\textup{(CLIP4Clip)-tightTransf} \dag$ \cite{luo2022clip4clip} & 26065.6 & 20.3 & 49.2 & 65.1 & 6.0 & 26.4 & 134.6 & 19.2 & 49.6 & 65.2 & 6.0 & 22.0 & 134.0 \\
			$\textup{(Ours)-tightTransf + Expansion Excitation}$ & 26102.7 & \textbf{21.4} & \textbf{51.1} & \textbf{65.8} & \textbf{5.0} & \textbf{23.0} & \textbf{138.3} & \textbf{22.0} & \textbf{51.4} & \textbf{67.1} & \textbf{5.0} & \textbf{20.7} & \textbf{140.5} \\
			\hline
		\end{tabular}
	}
\end{table*}

\section{Ablation Study}
\label{Ablation}
In this section, we conduct an ablation study on the MSR-VTT dataset to demonstrate the advantage of our proposed modules. We also design ablation experiments in the squeeze excitation-and-aggregation module to investigate the impact of different reduction ratios. 

\subsection{Evaluation on Squeeze Excitation-and-Aggregation Module} 
\label{sec:Squeeze Excitation-and-Aggregation Module}
To fully justify the effectiveness of squeeze excitation-and-aggregation module, we compare the performance of this hybrid module with three frame aggregation methods, including excitation module plus mean pooling, aggregation module, and excitation-and-aggregation module in different combinations. As can be seen from Table \ref{table: ablation study on meanP}, we gain three important observations.

\begin{figure}[!t]
	\centerline{\includegraphics[width=0.48\textwidth]{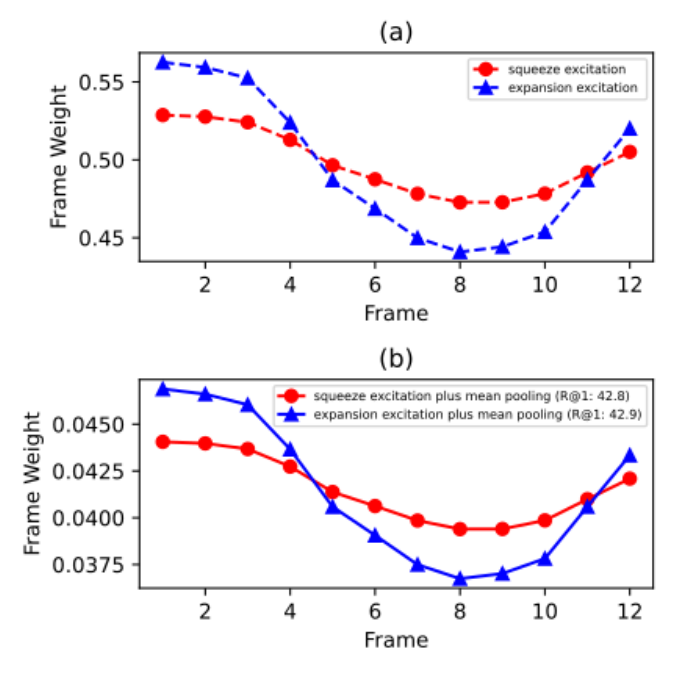}}
	\caption{(a) The distribution of frame-wise weights in squeeze / expansion excitation modules. (b) The distribution of frame-wise weights in squeeze / expansion excitation modules plus mean pooling. }  
	\label{fig:Comparison_of_frame_level_weight_in_excitation_module}
\end{figure}

\subsubsection{Excitation module plus mean pooling} Against our expectation that the excitation modules learn frame-wise relationships and attain better performance through mean pooling implementation, inserting squeeze excitation module achieves an lower R@1 of 42.8 with 0.7\% (0.3\%) relative (absolute) decline. We suppose this may be attributed to the small frame-wise weights activated by excitation module plus mean pooling, and we present the distribution visualization of frame-wise weights in Fig. \ref{fig:Comparison_of_frame_level_weight_in_excitation_module}, where the dashed red and blue lines in the subplot (a) are weights activated by the excitation modules in squeeze and expansion paradigms while the excitation weights through mean pooling are shown in the subplot (b). Surprisingly, the activated weights brought by excitation module plus mean pooling are about half of the mean pooling aggregation weights of 0.0833, which hinders the utilization of frame features and aggravates the smoothness of frame representation distribution. Moreover, we observe that the expansion module achieves 42.9 R@1 with +0.1\% absolute improvement over squeeze module, demonstrating the superiority of a wider range in frame-wise weight distribution. Most notably, compared with the proposed squeeze excitation-and-aggregation module, the squeeze excitation plus mean pooling is an inferior alternative, which decreases the performance of R@1 and RSum by 0.7\% and 2.1\%.

\begin{table*}
	\caption{Retrieval performance comparison on the DiDeMo dataset. Speed is the inference time per video during evaluation on a NVIDIA GeForce RTX 3090 GPU. $\dag$ indicates re-training.}
	\label{table: DiDeMo Performance}
	\centering
	\resizebox{\textwidth}{!}{
		\renewcommand{\arraystretch}{1.05} 
		\begin{tabular}{l|c|cccccc|cccccc}
			\hline \rule{0pt}{8pt}
			\multirow{2}*{Methods} & \multirow{2}*{\makecell[c]{Speed\\(ms)}} & \multicolumn{6}{c|}{Text-to-Video} &  \multicolumn{6}{c}{Video-to-Text} \\ 
			\cline{3-8} \cline{9-14}  
			&& R@1$\uparrow$  & R@5$\uparrow$  & R@10$\uparrow$ & MdR$\downarrow$ &  MnR$\downarrow$ & 
			RSum$\uparrow$
			& R@1$\uparrow$  & R@5$\uparrow$  & R@10$\uparrow$ & MdR$\downarrow$ &  MnR$\downarrow$ & 
			RSum$\uparrow$ \\
			\hline
			$\textup{S2VT}$\cite{venugopalan2014translating} & - & 11.9 & 33.6 & - & 13.0 & - & - & 13.2 & 33.6 & - & 15.0 & - & - \\
			$\textup{FSE}$\cite{zhang2018cross} & - & 13.9 & 36.0 & - & 11.0 & - & - & 13.1 & 33.9 & - & 12.0 & - & - \\
			$\textup{CE}$\cite{liu2019use} & - & 16.1 & 41.1 & - & 8.3 & 43.7 & - & 15.6 & 40.9 & - & 8.2 & 42.4 & - \\ 
			$\textup{Frozen}$\cite{bain2021frozen} & - & 34.6 & 65.0 & 74.7 & 3.0 & - & 174.3 & - & - & - & - & - & - \\
			$\textup{CLIPBERT}$\cite{lei2021less} & - & 20.4 & 48.0 & 60.8 & 6.0 & - & 129.2 & - & - & - & - & - & - \\
			$\textup{TT-CE+}$\cite{croitoru2021teachtext} & - & 21.6 & 48.6 & 62.9 & 6.0 & - & 133.1 & 21.1 & 47.3 & 61.1 & 6.3 & - & 129.5 \\
			\hline
			$\textup{(CLIP4Clip)-meanP} \dag$ & 18.8 & \textbf{40.8} & 67.8 & \textbf{77.4} & \textbf{2.0} & \textbf{20.8} & \textbf{186.0} & 40.7 & \textbf{67.3} & 77.5 & \textbf{2.0} & \textbf{15.4} & 185.5 \\
			$\textup{(Ours)-Expansion Excitation-and-Aggregation}$ & 29.6 & 40.1 & \textbf{67.9} & 75.9 & \textbf{2.0} & 20.9 & 183.9 & \textbf{41.0} & \textbf{67.3} & \textbf{77.6} & \textbf{2.0} & 18.3 & \textbf{185.9} \\
			\hline
			$\textup{(CLIP4Clip)-seqLSTM}\dag$ & 65.9 & \textbf{40.8} & 67.8 & \textbf{77.1} & \textbf{2.0} & \textbf{21.5} & \textbf{185.7} & 40.1 & 66.8 & \textbf{77.4} & \textbf{2.0} & \textbf{15.7} & 184.3 \\
			$\textup{(Ours)-seqLSTM + Expansion Aggregation}$ & 135.2 & 38.8 & \textbf{68.3} & 75.8 & \textbf{2.0} & 22.0 & 182.9 & \textbf{40.7} & \textbf{67.1} & 77.1 & \textbf{2.0} & 18.2 & \textbf{184.9} \\
			\hline
			$\textup{(CLIP4Clip)-seqTransf}\dag$ & 198.2 & \textbf{40.6} & \textbf{66.3} & \textbf{76.1} & \textbf{2.0} & \textbf{19.8} & \textbf{183.0} & \textbf{40.5} & 65.6 & 75.9 & \textbf{2.0} & \textbf{15.5} & \textbf{182.0} \\
			$\textup{(Ours)-seqTransf + Expansion Aggregation}$ & 252.0 & 39.2 & 65.7 & 75.7 & \textbf{2.0} & 20.1 & 180.6 & 39.3 & \textbf{65.9} & \textbf{76.5} & \textbf{2.0} & 16.2 & 181.7 \\
			\hline
			$\textup{(CLIP4Clip)-tightTransf}\dag$ & 5008.1 & 28.3 & 58.3 & 70.4 & 4.0 & 25.1 & 157.0 & 32.5 & 58.8 & 70.4 & 4.0 & 19.1 & 161.7 \\
			$\textup{(Ours)-tightTransf + Expansion Excitation}$ & 5026.5 & \textbf{31.0} & \textbf{59.7} & \textbf{71.8} & \textbf{3.0} & \textbf{23.8} & \textbf{162.5} & \textbf{34.6} & \textbf{61.2} & \textbf{70.5} & \textbf{3.0} & \textbf{17.2} & \textbf{166.3} \\
			\hline
		\end{tabular}
	}
\end{table*}

\begin{table*}
	\caption{Retrieval performance with different frame aggregation methods on the MSR-VTT dataset. ``meanP'' is short for mean pooling. }
	\label{table: ablation study on meanP}
	\centering
	\resizebox{\textwidth}{!}{
		\renewcommand{\arraystretch}{1.05} 
		\begin{tabular}{l|cccccc|cccccc}
			\hline
			\multirow{2}*{Methods} &  \multicolumn{6}{c|}{Text-to-Video} & \multicolumn{6}{c}{Video-to-Text} \\ 
			\cline{2-7} \cline{8-13} 
			& R@1$\uparrow$  & R@5$\uparrow$  & R@10$\uparrow$ & MdR$\downarrow$ &  MnR$\downarrow$ & RSum$\uparrow$
			& R@1$\uparrow$ & R@5$\uparrow$  & R@10$\uparrow$ & MdR$\downarrow$ &  MnR$\downarrow$ &
			RSum$\uparrow$ \\
			\hline
			meanP & 43.1 & 70.8 & 80.6 & 2.0 & 17.0 & 194.5 & 42.6 & 70.0 & 80.6 & 2.0 & 12.1 & 193.2 \\
			\hline
			squeeze excitation + meanP & 42.8 & 70.2 & 80.5 & \textbf{2.0} & \textbf{17.4} & 193.5 & 41.7 & 68.8 & \textbf{79.5} & \textbf{2.0} & 12.5 & 190.0 \\
			expansion excitation + meanP & \textbf{42.9} & \textbf{71.3} & \textbf{80.8} & \textbf{2.0} & \textbf{17.4} & \textbf{195.0} & \textbf{42.0} & \textbf{68.9} & \textbf{79.5} & \textbf{2.0} & \textbf{12.4} & \textbf{190.4} \\ 
			\hline
			squeeze aggregation& 43.2 & 70.8 & \textbf{79.9} & \textbf{2.0} & \textbf{16.7} & 193.9 & \textbf{42.7} & \textbf{70.5} & \textbf{80.5} & \textbf{2.0} & \textbf{11.9} & \textbf{193.7} \\
			expansion aggregation & \textbf{43.3} & \textbf{71.0} & \textbf{79.9} & \textbf{2.0} & 17.1 & \textbf{194.2} & 41.9 & 69.5 & 79.2 & \textbf{2.0} & 12.3 & 190.6 \\  
			\hline
			squeeze excitation + squeeze aggregation & \textbf{43.5} & \textbf{71.1} & \textbf{81.0} & \textbf{2.0} & 15.7 & \textbf{195.6} & 42.3 & 70.8 & 80.9 & \textbf{2.0} & 11.7 & 194.0 \\
			expansion excitation + expansion aggregation & \textbf{43.5} & 70.9 & 80.2 & \textbf{2.0} & 16.6 & 194.6 & \textbf{42.9} & 70.7 & 81.1 & \textbf{2.0} & 11.7 & \textbf{194.7} \\ 
			squeeze excitation + expansion aggregation & 43.1 & \textbf{71.1} & 80.4 & \textbf{2.0} & \textbf{15.6} & 194.6 & 41.9 & \textbf{71.0} & \textbf{81.4} & \textbf{2.0} & \textbf{11.4} & 194.3 \\ 
			expansion excitation + squeeze aggregation & 43.2 & 70.9 & 80.1 & \textbf{2.0} & 17.0 & 194.2 & 41.5 & 68.9 & 79.2 & \textbf{2.0} & 12.3 & 189.6 \\ 
			\hline
		\end{tabular}
	}
\end{table*}

\subsubsection{Aggregation Module} 
\begin{figure}[!t]
	\centerline{\includegraphics[width=0.48\textwidth]{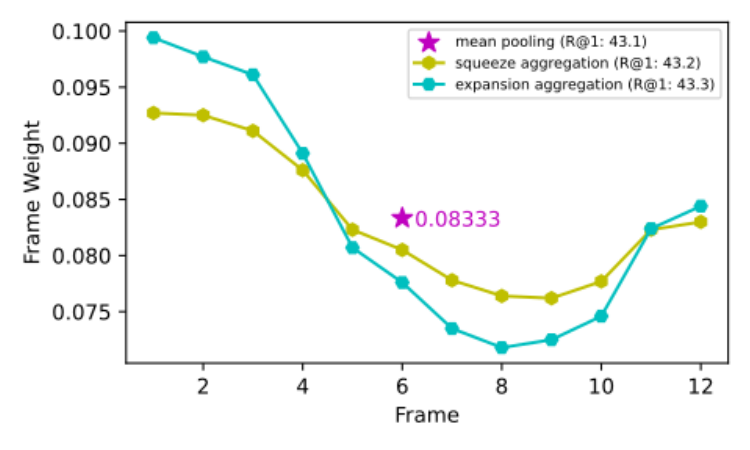}}
	\caption{The distribution of frame-wise weights in squeeze / expansion aggregation modules. The expansion aggregation module outperforms squeeze aggregation and pure mean pooling by +0.1\% and +0.2\% absolute improvements respectively. }
	\label{fig:Comparison_of_frame_level_weight_in_aggregation_module}
\end{figure}

The 43.3 R@1 indicates the strength of expansion aggregation module, which brings +0.5\% (+0.2\%) relative (absolute) improvements. To conduct quantitative analysis, we similarly  visualize the distribution of frame-wise weights in Fig. \ref{fig:Comparison_of_frame_level_weight_in_aggregation_module}. Unlike the narrow magnitude of frame-wise weights in the excitation modules, the aggregation modules generate weights in a larger magnitude, which is beneficial for frame feature utilization. It is noticeable that the mutual exclusion among frame-wise weights brought by aggregation modules matches with the mutually exclusive character of frame features, thus achieving better retrieval performance. The performance comparison (43.2 R@1 \textit{vs.} 43.3 R@1) between squeeze and expansion aggregation modules also verifies the essence of a wider range distribution of frame-wise weights. Besides, We also observe that different from the best R@1 of 43.5 provided by the squeeze excitation-and-aggregation module, the pure adoption of aggregation module brings lower R@1 around 43.2, which reveals the significance of excitation module in non-mutually-exclusive relationships capture. 

\subsubsection{Excitation-and-Aggregation Module} For simplicity, we first define excitation and aggregation modules in the same type as unified paradigms, such as squeeze excitation plus squeeze aggregation (squeeze \& squeeze) and expansion excitation plus expansion aggregation (expansion \& expansion). The remaining two schemes are divided into reversed paradigms, including squeeze excitation plus expansion aggregation (squeeze \& expansion) and expansion excitation plus squeeze aggregation (expansion \& squeeze). 

\begin{figure*}[!t]
	\centerline{\includegraphics[width=\textwidth]{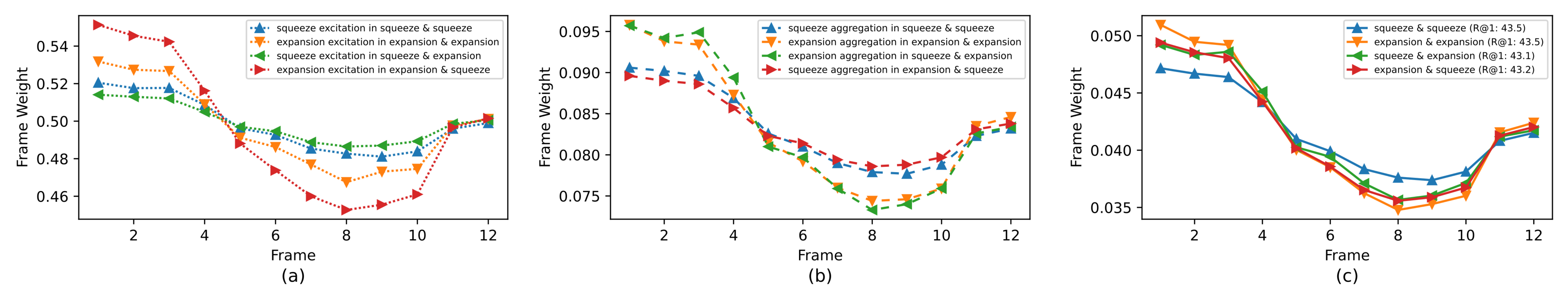}}
	\caption{(a) The distribution of frame-wise weights in pure squeeze / expansion excitation modules. (b) The distribution of frame-wise weights in pure squeeze / expansion aggregation modules. (c) The distribution of frame-wise weights in hybrid of squeeze / expansion excitation and aggregation modules. }
	\label{fig:Comparison_of_frame_level_weight_in_excitation_aggregation_module}
\end{figure*}

Then, the retrieval results are given in Table \ref{table: ablation study on meanP}. From the table, we can see that the unified paradigms achieve better results of 43.5 R@1 than the reversed paradigms, bringing +0.4\% absolute performance improvement. It is also worth noting that the squeeze excitation-and-aggregation module obtains the best RSum of 195.6, which further verifies the strength of the unified paradigms. Fig. \ref{fig:Comparison_of_frame_level_weight_in_excitation_aggregation_module} shows the frame-wise weights distribution visualization of excitation module, aggregation module, and excitation-and-aggregation module respectively. Compared with the consistent frame-wise weights distribution in the unified paradigms, the reversed paradigms present extreme fluctuation ranges, which is harmful to the retrieval performance, thus leading to about 0.4\% absolute decrease. 

\begin{table*}
	\caption{Retrieval performance with different frame aggregation methods in the seqLSTM type on the MSR-VTT dataset. ``seqLSTM'' is short for sequential LSTM. }
	\label{table: ablation study in seqLSTM}
	\centering
	\renewcommand{\arraystretch}{1.1}
	\resizebox{\textwidth}{!}{
		\renewcommand{\arraystretch}{1.05} 
		\begin{tabular}{l|cccccc|cccccc}
			\hline
			\multirow{2}*{Methods} &  \multicolumn{6}{c|}{Text-to-Video} & \multicolumn{6}{c}{Video-to-Text} \\ 
			\cline{2-7} \cline{8-13} 
			& R@1$\uparrow$  & R@5$\uparrow$  & R@10$\uparrow$ & MdR$\downarrow$ &  MnR$\downarrow$ & RSum$\uparrow$
			& R@1$\uparrow$ & R@5$\uparrow$  & R@10$\uparrow$ & MdR$\downarrow$ &  MnR$\downarrow$ & RSum$\uparrow$\\
			\hline
			seqLSTM & 42.4 & 69.4 & 80.1 & 2.0 & 17.2 & 191.9 & 40.9 & 69.8 & 79.5 & 2.0 & 13.0 & 190.2 \\
			\hline
			seqLSTM + squeeze excitation + meanP & 42.8 & \textbf{71.0} & 79.5 & \textbf{2.0} & 17.2 & 193.3 & 41.3 & \textbf{70.3} & 80.1 & \textbf{2.0} & 12.4 & 191.7 \\
			seqLSTM + expansion excitation + meanP & \textbf{43.7} & 70.1 & \textbf{81.0} & \textbf{2.0} & \textbf{16.7} & \textbf{194.8} & \textbf{42.6} & 69.5 & \textbf{80.3} & \textbf{2.0} & \textbf{12.0} & \textbf{192.4} \\ 
			\hline
			seqLSTM + squeeze aggregation & 43.1 & \textbf{71.1} & 80.2 & \textbf{2.0} & \textbf{16.4} & 194.4 & \textbf{42.9} & \textbf{70.3} & 80.9 & \textbf{2.0} & 12.0 & \textbf{194.1} \\
			seqLSTM + expansion aggregation & \textbf{43.5} & 70.5 & \textbf{80.8} & \textbf{2.0} & \textbf{16.4} & \textbf{194.8} & 42.2 & 69.7 & \textbf{81.2} & \textbf{2.0} & \textbf{11.9} & 193.1 \\  
			\hline
			seqLSTM + squeeze excitation + squeeze aggregation & \textbf{42.5} & 70.4 & 80.1 & \textbf{2.0} & 16.8 & 193.0 & 41.4 & 69.7 & 79.8 & \textbf{2.0} & 12.4 & 190.9 \\
			seqLSTM +  expansion excitation + expansion aggregation & 42.4 & 70.9 & 80.3 & \textbf{2.0} & 16.8 & 193.6 & \textbf{42.4} & 70.3 & 80.7 & \textbf{2.0} & 12.1 & 193.4 \\ 
			seqLSTM + squeeze excitation + expansion aggregation & \textbf{42.5} & \textbf{71.3} & \textbf{80.5} & \textbf{2.0} & \textbf{15.8} & \textbf{194.3} & 42.2 & \textbf{70.9} & \textbf{80.8} & \textbf{2.0} & \textbf{11.8} & \textbf{193.9} \\ 
			seqLSTM + expansion excitation + squeeze aggregation & 42.4 & 70.4 & 79.6 & \textbf{2.0} & 16.4 & 192.4 & 41.5 & 69.6 & 79.2 & \textbf{2.0} & 12.3 & 190.3 \\ 
			\hline
		\end{tabular}
	}
\end{table*}

\subsection{Evaluation on Expansion Aggregation Module in Sequential LSTM} 
\label{sec: Sequential LSTM}
To explore the impact of expansion aggregation module in sequential LSTM, we conduct an ablative study to compare our proposed module with other frame aggregation methods, like excitation module plus mean pooling, aggregation module, and excitation-and-aggregation module. From the results in Table \ref{table: ablation study in seqLSTM}, we give detail analysis from three aspects.

\subsubsection{Excitation Module plus Mean Pooling} 
\begin{figure}[!t]
	\centerline{\includegraphics[width=0.48\textwidth]{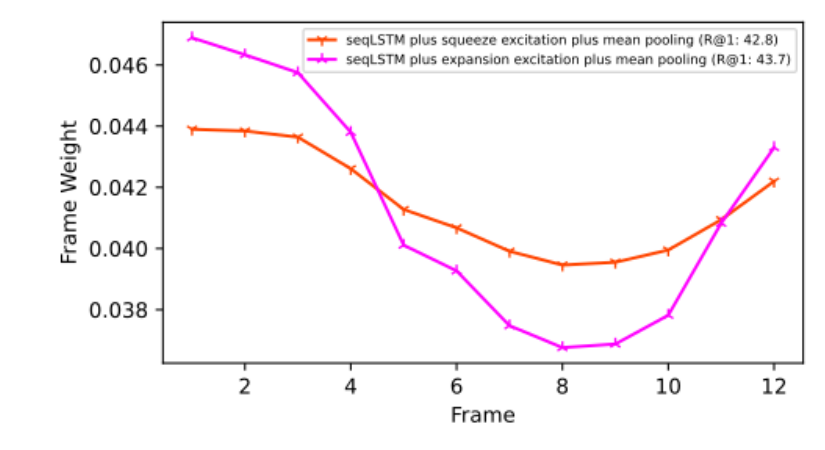}}
	\caption{The distribution of frame-wise weights in seqLSTM plus squeeze / expansion excitation modules plus mean pooling. In the sequential LSTM type, the R@1 of expansion excitation module outperforms squeeze excitation module by +0.9\% absolute improvement. }  
	\label{fig:Comparison_of_frame_level_weight_in_seqLSTM_excitation_module}
\end{figure}

Contrary to the performance decline from simple mean pooling to a combination of excitation modules plus mean pooling, we observe significant performance improvements from the fusion of sequential LSTM and excitation module plus mean pooling. Most precisely, the expansion excitation module achieves 43.7 R@1 and 194.8 RSum, bringing +3.1\% (+1.3\%) and +1.5\% (+2.9\%) relative (absolute) performance improvements over simple sequential LSTM plus mean pooling. The visualization of  frame-wise weights distribution is shown in Fig. \ref{fig:Comparison_of_frame_level_weight_in_seqLSTM_excitation_module}, where the weights are targeted at frame features output from sequential LSTM. We observe that although the generated weights are still half of the mean pooling aggregation weights, the retrieval performance has increased instead of a subtle decline. The opposite trend may be related to the temporal module, which 
has assigned different weights to distinct frame features, and some weights are larger than 1. This phenomenon  enlarges the subtle discrepancies among multiple frames and enhances the utilization of several critical frames, thus boosting the performance by a large margin. Furthermore, compared with the 42.8 R@1 brought by the squeeze excitation module, the expansion excitation module achieves better R@1 of 43.7 with +0.9\% absolute improvement, proving the superiority of a wider range of frame-wise weights distribution. It is worth noting that due to the lower RSum of 192.4 in the video-text retrieval task, the excitation module plus mean pooling is inferior to the proposed expansion aggregation module, which achieves 193.1 RSum with +0.7\% absolute improvement.

\subsubsection{Aggregation Module} 
\begin{figure}[!t]
	\centerline{\includegraphics[width=0.48\textwidth]{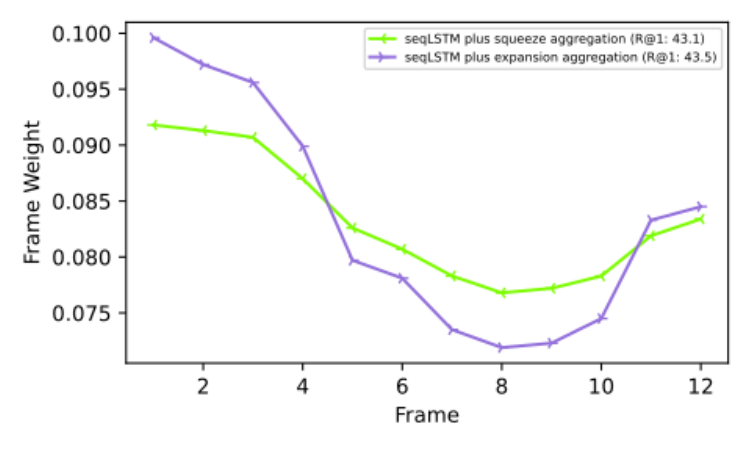}}
	\caption{The distribution of frame-wise weights in seqLSTM plus squeeze / expansion aggregation modules. The expansion aggregation module has the advantage over squeeze aggregation module, bringing +0.4\% absolute improvement. }
	\label{fig:Comparison_of_frame_level_weight_in_seqLSTM_aggregation_module}
\end{figure}

As shown in Table \ref{table: ablation study in seqLSTM}, we observe that based on the simple sequential LSTM, both squeeze and expansion aggregation modules attain better results. Specifically, the squeeze aggregation module achieves 43.1 R@1 and 194.4 RSum, outperforming the results in pure sequential LSTM by +1.7\% (+0.7\%) and +1.3\% (+2.5\%) relative (absolute) improvements. When the expansion aggregation module is utilized in the frame features aggregation, the R@1 of 43.5 and RSum of 194.8 are obtained, bringing +0.4\% absolute improvement. Additionally, from the visualization of frame-weights distribution in Fig. \ref{fig:Comparison_of_frame_level_weight_in_seqLSTM_aggregation_module}, we conclude that a wider range of frame-wise weights distribution is essential for retrieval performance gain.

\subsubsection{Excitation-and-Aggregation Module} There are three important observations from the comprehensive analysis of results in Table \ref{table: ablation study in seqLSTM}.  

$\bullet$ The combination of excitation and aggregation modules does not bring significant performance boost. We suppose that LSTM is insufficient for the full utilization of strength in the excitation-and-aggregation module, thus achieving the almost identical R@1 of 42.5 compared to the 42.4 R@1 in the pure sequential LSTM.

$\bullet$ The squeeze $\&$ expansion obtains the best performance, whereas the expansion $\&$ squeeze achieves the worst results. We think that the sequential module may be compatible with the reversed paradigm instead of the unified paradigm, and the slight temporal modeling capability brought by LSTM makes one reversed paradigm (42.5 R@1 in squeeze \& expansion) perform better than the unified paradigm (42.4 R@1 in expansion \& expansion). 

$\bullet$ Compared with the obtained 43.5 R@1 in the expansion aggregation module, introducing the expansion excitation module achieves 42.4 R@1 with 1.1\% absolute performance decrease, verifying the weakness of excitation module in the combination of sequential LSTM and hybrid modules.

\subsection{Evaluation on Expansion Aggregation Module in Sequential Transformer} 
\label{sec: Sequential Transformer}
To examine the impact of expansion aggregation module in sequential transformer encoder, we design an ablation study to compare our method with three frame aggregation methods, termed excitation module plus mean pooling, aggregation module and excitation-and-aggregation module respectively. Based on the ablation experimental results presented in Table \ref{table: ablation study in seqTransf}, we give three comprehensive analysis. 

\begin{table*}
	\caption{Retrieval performance with different frame aggregation methods in the seqTransf type on the MSR-VTT dataset. ``seqTransf'' is short for sequential Transformer Encoder.}
	\label{table: ablation study in seqTransf}
	\centering
	\renewcommand{\arraystretch}{1.1}
	\resizebox{\textwidth}{!}{
		\renewcommand{\arraystretch}{1.05} 
		\begin{tabular}{l|cccccc|cccccc}
			\hline
			\multirow{2}*{Methods} &  \multicolumn{6}{c|}{Text-to-Video} & \multicolumn{6}{c}{Video-to-Text} \\ 
			\cline{2-7} \cline{8-13}
			& R@1$\uparrow$  & R@5$\uparrow$  & R@10$\uparrow$ & MdR$\downarrow$ &  MnR$\downarrow$ & RSum$\uparrow$
			& R@1$\uparrow$ & R@5$\uparrow$  & R@10$\uparrow$ & MdR$\downarrow$ &  MnR$\downarrow$ & RSum$\uparrow$\\
			\hline
			seqTransf & 43.4 & 71.8 & 81.3 & 2.0 & 16.4 & 196.5 & 44.1 & 70.6 & 80.0 & 2.0 & 12.0 & 194.7 \\
			\hline
			seqTransf + squeeze excitation + meanP & \textbf{43.8} & \textbf{71.7} & \textbf{81.8} & \textbf{2.0} & 16.2 & \textbf{197.3} & \textbf{43.8} & \textbf{70.8} & 80.4 & \textbf{2.0} & 11.9 & \textbf{195.0} \\
			seqTransf +  expansion excitation + meanP & 43.7 & 70.8 & 81.6 & \textbf{2.0} & \textbf{16.1} & 196.1 & 43.2 & 69.1 & \textbf{80.6} & \textbf{2.0} & \textbf{11.8} & 192.9 \\ 
			\hline
			seqTransf + squeeze aggregation & 43.5 & 71.1 & 81.1 & \textbf{2.0} & 16.0 & 195.7 & \textbf{44.0} & 69.2 & \textbf{80.3} & \textbf{2.0} & 11.9 & 193.5 \\
			seqTransf + expansion aggregation & \textbf{43.9} & \textbf{71.2} & \textbf{82.0} & \textbf{2.0} & \textbf{15.9} & \textbf{197.1} & 43.7 & \textbf{69.9} & \textbf{80.3} & \textbf{2.0} & \textbf{11.8} & \textbf{193.9} \\  
			\hline
			seqTransf + squeeze excitation + squeeze aggregation & 43.3 & 71.6 & 81.4 & \textbf{2.0} & 16.0 & 196.3 & 43.2 & 69.3 & 80.4 & \textbf{2.0} & 12.0 & 192.9 \\
			seqTransf + expansion excitation + expansion aggregation & 43.1 & 71.1 & 81.2 & \textbf{2.0} & \textbf{15.8} & 195.4 & \textbf{43.9} & 69.5 & 80.8 & \textbf{2.0} & \textbf{11.7} & 194.2 \\ 
			seqTransf + squeeze excitation + expansion aggregation & \textbf{43.7} & \textbf{71.8} & \textbf{82.1} & \textbf{2.0} & \textbf{15.8} & \textbf{197.6} & 43.8 & \textbf{70.4} & \textbf{81.0} & \textbf{2.0} & \textbf{11.7} & \textbf{195.2} \\ 
			seqTransf + expansion excitation + squeeze aggregation & \textbf{43.7} & 71.3 & 81.1 & \textbf{2.0} & 15.9 & 196.1 & 43.6 & 69.7 & 80.5 & \textbf{2.0} & 11.8 & 193.8 \\ 
			\hline
		\end{tabular}
	}
\end{table*}

\subsubsection{Excitation Module plus Mean Pooling} 
\begin{figure}[!t]
	\centerline{\includegraphics[width=0.48\textwidth]{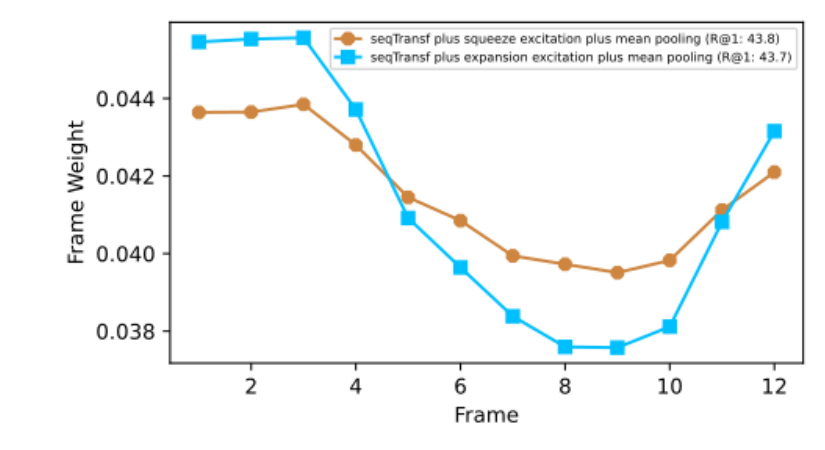}}
	\caption{The distribution of frame-wise weights in seqTransf plus squeeze / expansion excitation modules plus mean pooling. The squeeze excitation module achieves 43.8 R@1, outperforming the results in the expansion excitation module by +0.1\% absolute improvement. }  
	\label{fig:Comparison_of_frame_level_weight_in_seqTransf_excitation_module}
\end{figure}

Compared with the single transformer, the squeeze and expansion excitation modules achieve 43.8 R@1 (+0.4\% absolute improvement) and 43.7 R@1 (+0.3\% absolute improvement) respectively, further demonstrating the effectiveness of excitation modules in the sequential type. Fig. \ref{fig:Comparison_of_frame_level_weight_in_seqTransf_excitation_module} mainly illustrates the distribution of frame-wise weights, which are targeted at the output from transformer encoder. Contrary to the the results of excitation modules in the sequential LSTM, squeeze excitation is superior to the expansion excitation with +0.1\% absolute performance improvement. We suppose that the performance discrepancy is relevant to the temporal modeling ability. More precisely, the LSTM is a weaker sequential module while the transformer encoder possesses powerful temporal modeling ability. For the sequential LSTM, expansion excitation module introduces more nonlinearity to supplement with weaker temporal modeling ability, thus achieving better retrieval performance. However, the sequential transformer encoder is powerful enough and squeeze excitation module may be a better alternative. Besides, R@1 in the text-video retrieval task obtained from expansion excitation and expansion aggregation modules are 43.7 R@1 and 43.9 R@1 respectively, of which the +0.2\% absolute performance improvement justifies the advantage of expansion aggregation module.

\begin{figure}[!t]
	\centerline{\includegraphics[width=0.48\textwidth]{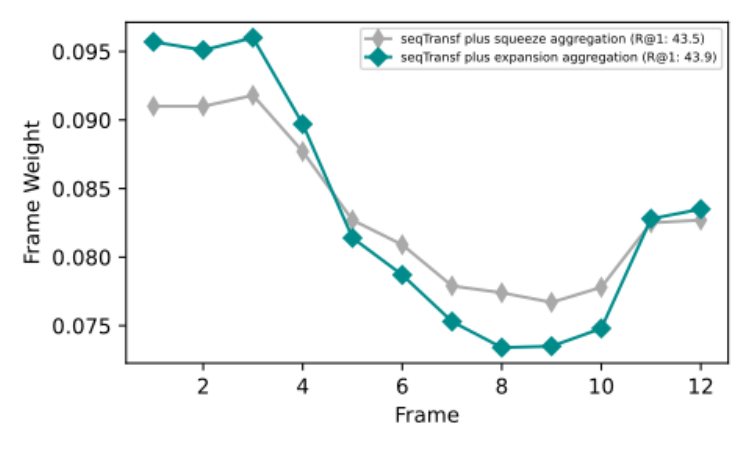}}
	\caption{The distribution of frame-wise weights in seqTransf plus squeeze / expansion aggregation modules. Compared with the squeeze aggregation module, expansion aggregation module is a better alternative, bringing +0.4\% absolute improvement. }
	\label{fig:Comparison_of_frame_level_weight_in_seqTransf_aggregation_module}
\end{figure}

\subsubsection{Aggregation Module} As reported in Table \ref{table: ablation study in seqTransf}, the expansion aggregation module achieves 43.9 R@1 and 197.1 RSum, surpassing the results in the pure sequential transformer and squeeze aggregation module by +0.5\%, +0.4\% and +0.6\%, +1.4\% absolute performance improvements. The visualization of frame-wise weights distribution in Fig. \ref{fig:Comparison_of_frame_level_weight_in_seqTransf_aggregation_module} satisfies our expectation that a wider range achieves better performance of 43.9 R@1 with +0.4\% absolute improvement. 

\subsubsection{Excitation-and-Aggregation module} The experimental results in Table \ref{table: ablation study in seqTransf} demonstrate the superiority of reversed paradigms over unified paradigms. Specifically, both squeeze $\&$ expansion and expansion $\&$ squeeze modules achieve 43.7 R@1, outperforming expansion \& expansion by +0.6\% absolute performance improvement. Different from the strength of squeeze $\&$ squeeze over expansion $\&$ squeeze in the sequential LSTM, the R@1 of expansion $\&$ squeeze surpasses squeeze $\&$ squeeze by +1.4\% relative improvement in the sequential Transformer. We think that the powerful sequential modeling ability brought by transformer matches with reversed paradigms, thus leading to the performance improvement over the unified paradigms.
Compared with the 43.9 R@1 obtained through the pure expansion aggregation module in the text-video retrieval task, the excitation-and-aggregation modules achieves the best R@1 of 43.7 with 0.2\% absolute decline, which verifies the weakness of excitation-and-aggregation modules.

\begin{table*}
	\caption{Retrieval performance with different frame aggregation methods in the tight type on the MSR-VTT dataset. ``tightTransf'' indicates Transformer Encoder for multi-modal interaction between video and text. }
	\label{table: ablation study on excitation}
	\centering
	\renewcommand{\arraystretch}{1.1}
	\resizebox{\textwidth}{!}{
		\renewcommand{\arraystretch}{1.05} 
		\begin{tabular}{l|cccccc|cccccc}
			\hline
			\multirow{2}*{Methods} &  \multicolumn{6}{c|}{Text-to-Video} & \multicolumn{6}{c}{Video-to-Text} \\ 
			\cline{2-7} \cline{8-13} 
			& R@1$\uparrow$  & R@5$\uparrow$  & R@10$\uparrow$ & MdR$\downarrow$ &  MnR$\downarrow$ & RSum$\uparrow$
			& R@1$\uparrow$ & R@5$\uparrow$  & R@10$\uparrow$ & MdR$\downarrow$ &  MnR$\downarrow$ & 
			RSum$\uparrow$ \\
			\hline
			tightTransf & 40.4 & 72.3 & 82.0 & 2.0 & 13.3 & 194.7 & 40.7 & 70.1 & 80.3 & 2.0 & 12.3 & 191.1 \\
			\hline
			tightTransf + squeeze excitation & \textbf{41.5} & 71.6 & \textbf{82.3} & \textbf{2.0} & \textbf{13.5} & \textbf{195.4} & 41.2 & \textbf{70.3} & \textbf{79.4} & \textbf{2.0} & \textbf{13.1} & \textbf{190.9} \\
			tightTransf + expansion excitation & \textbf{41.5} & \textbf{71.9} & 81.6 & \textbf{2.0} & 15.3 & 195.0 & \textbf{41.8} & 69.8 & 79.0 & \textbf{2.0} & 16.4 & 190.6 \\ 
			\hline
		\end{tabular}
	}
\end{table*}

\subsection{Evaluation on Squeeze Excitation Module in tight type}
\label{sec: Squeeze Excitation Module}
To examine the impact of the proposed squeeze excitation module, we conduct an ablation study to compare our method with expansion excitation module for frame features recalibration. From the experimental results in Table \ref{table: ablation study on excitation}, we observe that both excitation modules achieves better performance of 41.5 R@1 in the text-video retrieval task in comparison to the pure tight Transformer, bringing  +3.7\% (+1.1\%) relative (absolute) improvement. We also notice that although same R@1 are obtained through the squeeze and expansion excitation modules, the squeeze paradigm achieves better overall retrieval performance of 195.4 T2V RSum and 190.9 V2T RSum, outperforming the results in expansion paradigm by +0.4\% and +0.3\% absolute improvement. Moreover, lower MnR of 13.5 and 13.1 in the text-video and video-text tasks are obtained by the squeeze excitation module, which fully demonstrates the strength of squeeze excitation module in the tight type.

\begin{table*}
	\caption{Retrieval performance with squeeze excitation-and-aggregation module at different reduction ratios on the MSR-VTT dataset.}
	\label{table: different reduction ratios in squeeze excitation-and-aggregation module} 
	\centering
	\renewcommand{\arraystretch}{1.1}
	\resizebox{\textwidth}{!}{
		\renewcommand{\arraystretch}{1.05} 
		\begin{tabular}{c|cccccc|cccccc}
			\hline
			\multirow{2}*{Ratio $r$} & \multicolumn{6}{c|}{Text-to-Video} & \multicolumn{6}{c}{Video-to-Text} \\ 
			\cline{2-7} \cline{8-13} 
			& R@1$\uparrow$  & R@5$\uparrow$  & R@10$\uparrow$ & MdR$\downarrow$ &  MnR$\downarrow$ &
			RSum$\uparrow$
			& R@1$\uparrow$ & R@5$\uparrow$  & R@10$\uparrow$ & MdR$\downarrow$ &  MnR$\downarrow$ & 
			RSum$\uparrow$ \\
			\hline
			2 & 43.2 & \textbf{71.1} & 80.0 & \textbf{2.0} & 16.8 & 194.3 & \textbf{43.3} & 70.4 & \textbf{80.9} & \textbf{2.0} & 11.9 & \textbf{194.6} \\
			3 & \textbf{43.7} & \textbf{71.1} & 80.2 & \textbf{2.0} & 16.7 & 195.0 & 42.8 & 70.1 & 80.7 & \textbf{2.0} & 11.9 & 193.6 \\ 
			4 & 43.5 & \textbf{71.1} & \textbf{81.0} & \textbf{2.0} & \textbf{15.7} & \textbf{195.6} & 42.3 & \textbf{70.8} & \textbf{80.9} & \textbf{2.0} & \textbf{11.7} & 194.0 \\ 
			6 & 43.0 & 70.9 & 80.5 & \textbf{2.0} & 16.7 & 194.4 & 42.5 & 70.1 & 80.6 & \textbf{2.0} & \textbf{11.7} & 193.2 \\ 
			\hline
		\end{tabular}
	}
\end{table*}

\subsection{Effect of reduction ratio} 
The reduction ratio $r$ in our aforementioned squeeze-relevant modules defaults to 4. To investigate the impact  of this hyperparameter on retrieval performance, we conduct ablation experiments with squeeze excitation-and-aggregation module at a range of distinct $r$ values. As shown in Table \ref{table: different reduction ratios in squeeze excitation-and-aggregation module}, an increased ratio does not improve performance monotonically while a moderate reduction ratio of 4 is a better alternative. More specifically, setting $r=4$ achieves the best T2V RSum of 195.6, outperforming the results at other ratios by +0.7\% (+1.3\%), +0.3\% (+0.6\%) and +0.6\% (+1.2\%) relative (absolute) improvements. In practice, the reduction ratio must be carefully tuned to meet the requirement of cross-modal retrieval tasks.

\section{Conclusion}
\label{Conclusion}
In this paper, we study the drawbacks of mean pooling operation in frame representations aggregation and present several excitation and aggregation designs for CLIP4Clip retrieval results improvements. We first propose a novel excitation-and-aggregation design to encourage simultaneous modeling of non-mutually-exclusive and exclusive relationships. Then we introduce a new aggregation design in the sequential type, which aims to achieve temporal frame representations aggregation through the attention mechanism. We also improve the tight type with an excitation design to enhance multi-modal interaction and boost the comprehensive understanding of text-video contents. Experimental results on three benchmark datasets demonstrate the effectiveness of our proposed designs, achieving 43.9 R@1 (+1.2\% relative improvement, +0.5\% absolute improvement), 44.1 R@1 (+4.5\% relative improvement, +1.9\% absolute improvement), and 31.0 R@1 (+9.5\% relative improvement, +2.7\% absolute improvement) on the MSR-VTT, ActivityNet, and DiDeMo datasets. 

Although the proposed excitation and aggregation designs bring significant performance improvements in the video-text retrieval tasks, the direct utilization of all frame representations for discriminative video representation generation is laborious, especially in the video-paragraph paradigm with longer textual descriptions. In a human sense, not all frame features are semantic-relevant to the given text-based query. As a part of future work, we consider selecting the top-$k$ most semantically similar frames with higher similarity scores for subsequent aggregation operation while lower similarity frames are completely discarded, which can achieve a decent reduction in computational overhead and speed up the inference process. 
In addition, we plan on employing excitation and aggregation designs to other CLIP4Clip-based text-video retrieval models and other cross-modal tasks like video question answering in future research.

\section*{Acknowledgments}
This work was supported by the China National R\&D Key Research Program under Grants 2020YFB1711200 and 2019YFB1705700.


	\subsection*{  } 
	\setlength\intextsep{0pt} 
	\begin{wrapfigure}{l}{25mm}
		\centering
		\includegraphics[width=1in,height=1.25in,clip,keepaspectratio]{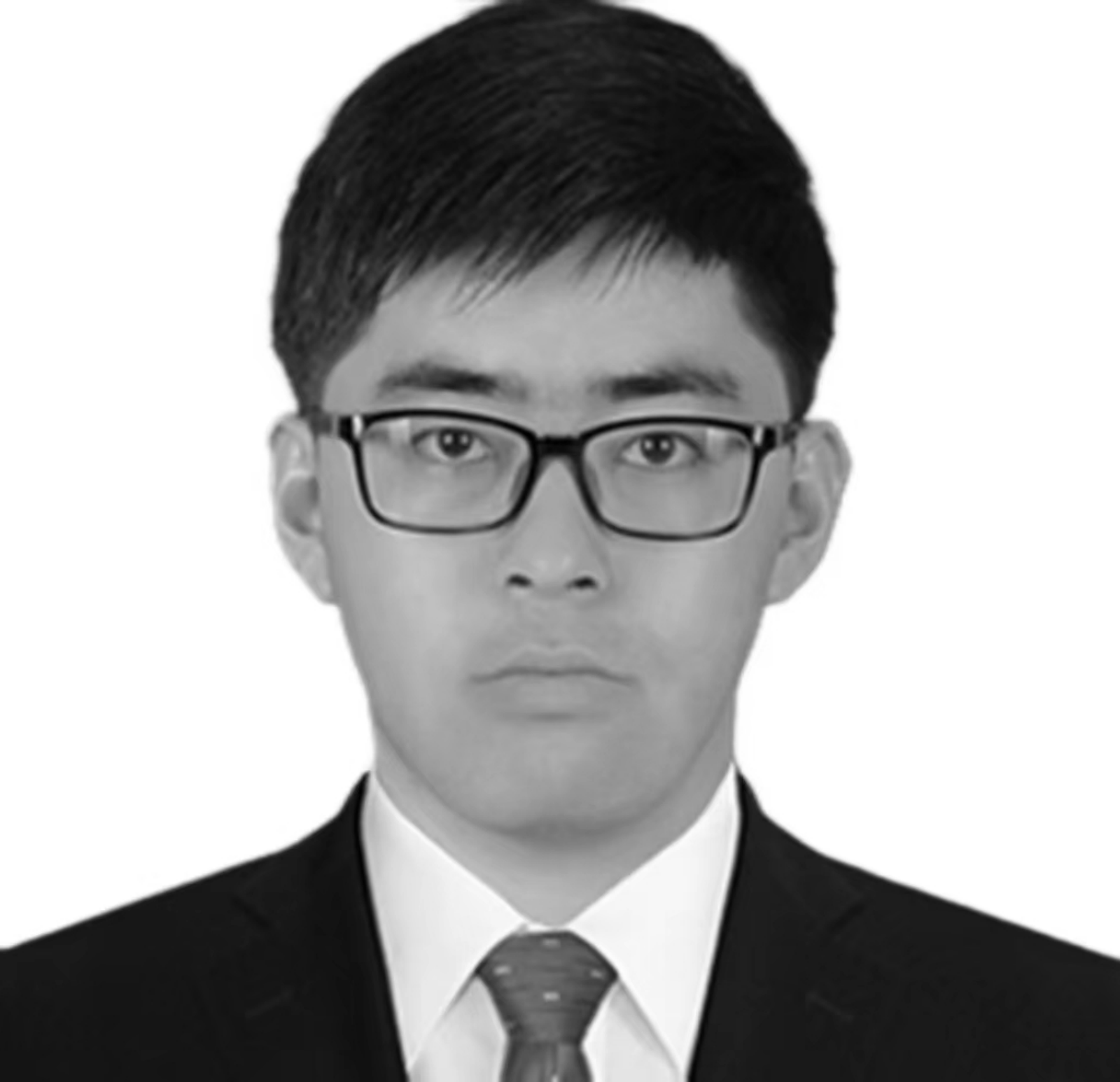}
	\end{wrapfigure}
	\noindent \textbf{Xiaolun Jing} received the B.S. degree from Central South University, in 2017, the M.S. degree from Xi'an Jiaotong University, in 2020. He is a Ph.D. candidate in the Department of Automation, Shanghai Jiao Tong University. His current research interests include video understanding, video action recognition and video-text retrieval in computer vision. \par
	
	\hspace*{\fill} 
	
	\subsection*{  } 
	\setlength\intextsep{0pt} 
	\begin{wrapfigure}{l}{25mm}
		\centering
		\includegraphics[width=1in,height=1.25in,clip,keepaspectratio]{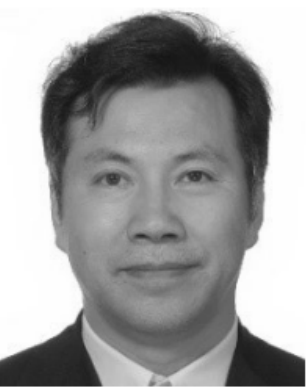}
	\end{wrapfigure}
	\noindent \textbf{Genke Yang} was born in Shanxi, China. He
	received the B.S. degree in mathematics from Shanxi University, in 1984, the M.S. degree in
	mathematics from Xi’nan Normal University, in 1987, and the Ph.D. degree in systems engineering from Xi’an Jiaotong University, in 1998. He has been a full-time Professor with the Department of Automation, Shanghai Jiao Tong University, Shanghai, China. He is currently a member of the Collaborative Innovation Center for Advanced Ship and Deep-Sea Exploration, Shanghai. His research interests include supply chain management, logistics, production planning and scheduling, discrete event dynamics systems, and computer integrated manufacturing. \par
	
	\hspace*{\fill} 
	
	\subsection*{  } 
	\setlength\intextsep{0pt} 
	\begin{wrapfigure}{l}{25mm}
		\centering
		\includegraphics[width=1in,height=1.25in,clip,keepaspectratio]{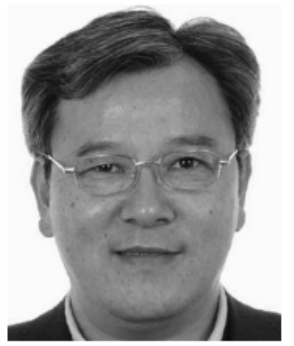}
	\end{wrapfigure}
	\noindent \textbf{Jian Chu} was born in 1963. He received the B.S., M.S., and Ph.D. degrees from Zhejiang University, Hangzhou, China, in 1982, 1984, and 1989, respectively, and the Ph.D. degree in Joint Education Program from Zhejiang University and Kyoto University, Kyoto, Japan. He was a Post-Doctoral Researcher with the Institute of Advanced Process Control, Zhejiang University, where he was a Full Professor in 1993, and a Doctorial Advisor in 1994. He is now the chief researcher of Shanghai Jiaotong University. His current research interests include control theory and applications, research and development of computer control systems, and advanced process control software. \par
	\enlargethispage{2cm}

\appendix
\onecolumn
\setcounter{table}{0}   
\setcounter{figure}{0}  
\setcounter{equation}{0} 

\renewcommand{\thetable}{A\arabic{table}}
\renewcommand{\thefigure}{A\arabic{figure}}
\renewcommand{\theequation}{A\arabic{equation}}

\section*{Appendix}
In this appendix, we give the details of performance analysis on the DiDeMo dataset in § \ref{app: performance comparison}. 

\begin{table*}\fontsize{3pt}{5pt}\selectfont
	\centering
	\caption{Data Statistics Comparison between the ActivityNet and DiDeMo datasets.}
	\setlength{\arrayrulewidth}{0.1pt}
	\label{table: data Statistics}
	\resizebox{\textwidth}{!}{
		\begin{tabular}{ccccc}
			\hline
			Datasets & Train Videos & Val Videos & Test Videos & Total Videos \\
			\hline
			ActivityNet & 10009 & 4917 & 4917 & 20k \\
			\hline
			DiDeMo & 8395 & 1065 & 1004 & 10k \\
			\hline
		\end{tabular}
	}
\end{table*}

\section{Performance Analysis on DiDeMo}
\label{app: performance comparison}
To clearly elaborate on the reasons behind the performance degradation on the DiDeMo dataset, we first select the ActivityNet dataset that also follows the video-paragraph retrieval paradigm as control, and give the data statistics comparison. As shown in Table \ref{table: data Statistics}, the video volume of DiDeMo is only about half of ActivityNet. Then we report the T2V retrieval results with different frame aggregation methods at three types on the ActivityNet and DiDeMo datasets in Table \ref{table: mean pooling type}, \ref{table: seqLSTM type} and \ref{table: seqTransf Type}, including mean pooling type, sequential LSTM type and sequential Transformer Encoder type.

\subsection{Mean Pooling Type}
As shown in Table \ref{table: mean pooling type}, we observe that compared with the squeeze paradigm, the expansion paradigm ia a better alternative in the video-paragraph retrieval tasks. More specifically, the cascade of expansion excitation module and mean pooling operation achieves 41.8 R@1 and 41.1 R@1 on the ActivityNet and DiDeMo datasets, outperforming the squeeze excitation results by +0.9\% (+0.4\%) and +0.2\% (+0.1\%) relative (absolute) improvements. When the aggregation module is utilized for frame aggregation, the expansion aggregation module also presents better RSum of 201.6 (\textit{i.e.} 4.3\% relative improvement, 8.3\% absolute improvement) and 185.2 (\textit{i.e.} 5.0\% relative improvement, 8.9\% absolute improvement) in the ActivityNet and DiDeMo retrieval tasks. We also notice that the expansion excitation-and-aggregation module boosts the squeeze paradigm R@1 by +6.8\% (+2.7\%) and +6.1\% (+2.3\%) relative (absolute) improvements on the ActivityNet and DiDeMo datasets. We think that the significant performance gain may be related to the video-paragraph retrieval paradigm, where the training of reasoning capability between video and longer captions needs more learnable parameters. Under this assumption, abundant trainable parameters introduced by the expansion paradigm can fully capture semantic relationships among different frames, thus promoting the retrieval performance by a large margin.

Surprisingly, we can see that DiDeMo retrieval results brought by the expansion excitation-and-aggregation module suffer a great decline, such as 0.7\% and 2.1\% absolute decrease at R@1 and RSum. The reason behind this degradation may be attributed to the incompatibility of training parameters and data volume. Generally, the mismatch between the amount of learnable parameters and the amount of data will lead to the risk of underfitting or overfitting. Compared to the 40.8 R@1 given by the mean pooling operation, the introduction of expansion excitation module brings additional parameters that match the data volume, thus promoting the R@1 by +0.7\% (+0.3\%) relative (absolute) improvements. However, the expansion excitation-and-aggregation module contains an excessive amount of learnable parameters that may overfit the training dataset, the model consequently achieves poor retrieval performance on the validation dataset. It is worth noting that this module achieves the best retrieval results on the ActivityNet dataset with larger data volume, which further verifies our speculation.

\subsection{Sequential LSTM Type}
In Table \ref{table: seqLSTM type}, we show the ActivityNet and DiDeMo retrieval results in the sequential LSTM type with different frame aggregation methods. Different from the better R@1 of 42.6 attained through the expansion aggregation module on the ActivityNet dataset, this module generates a lower R@1 of 38.8 with 2.0\% absolute decline. We explain the phenomenon difference based on the compatibility of training parameters and data volume. More precisely, the parameters contained in sequential LSTM and expansion aggregation module match with the data volume of ActivityNet, but seem to be overfitting on the DiDeMo dataset with smaller data volume, thus hindering the retrieval performance boost in the DiDeMo retrieval task.

\subsection{Sequential Transformer Encoder Type}
Table \ref{table: seqTransf Type} provides a comparison of the retrieval results between ActivityNet and DiDeMo in the sequential Transformer Encoder type with different frame aggregation methods. Similarly, we can see that although the expansion aggregation module boosts the ActivityNet R@1 by +4.5\% (+1.9\%) relative (absolute) improvement, it achieves 39.2 R@1 with 1.4\% absolute decrease on the DiDeMo dataset. The reason of performance decline results from the mismatch between the smaller data volume and the larger amount of learnable parameters contained in the cascade of sequential Transformer Encoder and expansion aggregation module, which increases the risk of overfitting and presents poor generalizability on unseen pairwise text-video data.

\begin{table*}[ht]
	\caption{T2V retrieval results with different frame aggregation methods on the ActivityNet and DiDeMo datasets. The number in blue is the retrieval result obtained by our mean pooling re-training.}
	\centering
	\renewcommand{\arraystretch}{1.1} 
	\label{table: mean pooling type}
	\resizebox{\textwidth}{!}{
		\begin{tabular}{l|cccccc|cccccc}
			\hline
			\multirow{2}*{Methods} &  \multicolumn{6}{c|}{ActivityNet Retrieval} & \multicolumn{6}{c}{DiDeMo Retrieval} \\ 
			\cline{2-7} \cline{8-13} 
			& R@1$\uparrow$  & R@5$\uparrow$  & R@10$\uparrow$ & MdR$\downarrow$ &  MnR$\downarrow$ & RSum$\uparrow$
			& R@1$\uparrow$ & R@5$\uparrow$  & R@10$\uparrow$ & MdR$\downarrow$ &  MnR$\downarrow$ &
			RSum$\uparrow$ \\
			\hline
			squeeze excitation + meanP & 41.4\textcolor{blue}{(41.1)} & 73.1\textcolor{blue}{(73.1)} & 84.7\textcolor{blue}{(84.6)} & \textbf{2.0}\textcolor{blue}{(2.0)} & 7.6\textcolor{blue}{(7.6)} & 199.2\textcolor{blue}{(198.8)} & 41.0\textcolor{blue}{(40.8)} & 67.8\textcolor{blue}{(67.8)} & \textbf{77.6}\textcolor{blue}{(77.4)} & \textbf{2.0}\textcolor{blue}{(2.0)} & 20.8\textcolor{blue}{(20.8)} & 186.4\textcolor{blue}{(186.0)} \\
			expansion excitation + meanP & \textbf{41.8}\textcolor{blue}{(41.1)} & \textbf{73.7}\textcolor{blue}{(73.1)} & \textbf{84.9}\textcolor{blue}{(84.6)} & \textbf{2.0}\textcolor{blue}{(2.0)} & \textbf{7.5}\textcolor{blue}{(7.6)} & \textbf{200.4}\textcolor{blue}{(198.8)} & \textbf{41.1}\textcolor{blue}{(40.8)} & \textbf{68.5}\textcolor{blue}{(67.8)} & 77.3\textcolor{blue}{(77.4)} & \textbf{2.0}\textcolor{blue}{(2.0)} & \textbf{20.6}\textcolor{blue}{(20.8)} & \textbf{186.9}\textcolor{blue}{(186.0)} \\ 
			\hline
			squeeze aggregation & 39.1\textcolor{blue}{(41.1)} & 71.3\textcolor{blue}{(73.1)} & 82.9\textcolor{blue}{(84.6)} & \textbf{2.0}\textcolor{blue}{(2.0)} & 11.3\textcolor{blue}{(7.6)} & 193.3\textcolor{blue}{(198.8)} & 37.3\textcolor{blue}{(40.8)} & 64.6\textcolor{blue}{(67.8)} & 74.4\textcolor{blue}{(77.4)} & 3.0\textcolor{blue}{(2.0)} & 30.6\textcolor{blue}{(20.8)} & 176.3\textcolor{blue}{(186.0)} \\
			expansion aggregation & \textbf{42.1}\textcolor{blue}{(41.1)} & \textbf{74.3}\textcolor{blue}{(73.1)} & \textbf{85.2}\textcolor{blue}{(84.6)} & \textbf{2.0}\textcolor{blue}{(2.0)} & \textbf{7.8}\textcolor{blue}{(7.6)} & \textbf{201.6}\textcolor{blue}{(198.8)} & \textbf{39.9}\textcolor{blue}{(40.8)} & \textbf{68.5}\textcolor{blue}{(67.8)} & \textbf{76.8}\textcolor{blue}{(77.4)} & \textbf{2.0}\textcolor{blue}{(2.0)} & \textbf{20.6}\textcolor{blue}{(20.8)} & \textbf{185.2}\textcolor{blue}{(186.0)} \\  
			\hline
			squeeze excitation + squeeze aggregation & 39.7\textcolor{blue}{(41.1)} & 72.0\textcolor{blue}{(73.1)} & 83.8\textcolor{blue}{(84.6)} & \textbf{2.0}\textcolor{blue}{(2.0)} & 10.9\textcolor{blue}{(7.6)} & 195.5\textcolor{blue}{(198.8)} & 37.8\textcolor{blue}{(40.8)} & 65.0\textcolor{blue}{(67.8)} & 74.5\textcolor{blue}{(77.4)} & \textbf{2.0}\textcolor{blue}{(2.0)} & 27.7\textcolor{blue}{(20.8)} & 177.3\textcolor{blue}{(186.0)} \\
			expansion excitation + expansion aggregation & \textbf{42.4}\textcolor{blue}{(41.1)} & \textbf{74.3}\textcolor{blue}{(73.1)} & \textbf{85.3}\textcolor{blue}{(84.6)} & \textbf{2.0}\textcolor{blue}{(2.0)} & \textbf{7.8}\textcolor{blue}{(7.6)} & \textbf{202.0}\textcolor{blue}{(198.8)} & 40.1\textcolor{blue}{(40.8)} & 67.9\textcolor{blue}{(67.8)} & \textbf{75.9}\textcolor{blue}{(77.4)} & \textbf{2.0}\textcolor{blue}{(2.0)} & \textbf{20.9}\textcolor{blue}{(20.8)} & 183.9\textcolor{blue}{(186.0)} \\ 
			squeeze excitation + expansion aggregation & 42.1\textcolor{blue}{(41.1)} & 74.2\textcolor{blue}{(73.1)} & 85.0\textcolor{blue}{(84.6)} & \textbf{2.0}\textcolor{blue}{(2.0)} & 8.0\textcolor{blue}{(7.6)} & 201.3\textcolor{blue}{(198.8)} & \textbf{40.4}\textcolor{blue}{(40.8)} & \textbf{68.3}\textcolor{blue}{(67.8)} & 75.6\textcolor{blue}{(77.4)} & \textbf{2.0}\textcolor{blue}{(2.0)} & 21.4\textcolor{blue}{(20.8)} & \textbf{184.3}\textcolor{blue}{(186.0)} \\ 
			expansion excitation + squeeze aggregation & 41.8\textcolor{blue}{(41.1)} & 73.8\textcolor{blue}{(73.1)} & 84.9\textcolor{blue}{(84.6)} & \textbf{2.0}\textcolor{blue}{(2.0)} & 8.6\textcolor{blue}{(7.6)} & 200.5\textcolor{blue}{(198.8)} & 39.9\textcolor{blue}{(40.8)} & 67.4\textcolor{blue}{(67.8)} & \textbf{75.9}\textcolor{blue}{(77.4)} & \textbf{2.0}\textcolor{blue}{(2.0)} & 22.7\textcolor{blue}{(20.8)} & 183.2\textcolor{blue}{(186.0)} \\ 
			\hline
		\end{tabular}
	}
\end{table*}

\begin{table*}[ht]
	\caption{T2V retrieval results with different frame aggregation methods in the seqLSTM type on the ActivityNet and DiDeMo datasets. The number in blue is the retrieval result obtained by our sequential LSTM re-training.}
	\centering
	\renewcommand{\arraystretch}{1.1} 
	\label{table: seqLSTM type}
	\resizebox{\textwidth}{!}{
		\begin{tabular}{l|cccccc|cccccc}
			\hline
			\multirow{2}*{Methods} &  \multicolumn{6}{c|}{ActivityNet Retrieval} & \multicolumn{6}{c}{DiDeMo Retrieval} \\ 
			\cline{2-7} \cline{8-13} 
			& R@1$\uparrow$  & R@5$\uparrow$  & R@10$\uparrow$ & MdR$\downarrow$ &  MnR$\downarrow$ & RSum$\uparrow$
			& R@1$\uparrow$ & R@5$\uparrow$  & R@10$\uparrow$ & MdR$\downarrow$ &  MnR$\downarrow$ &
			RSum$\uparrow$ \\
			\hline
			seqLSTM + squeeze excitation + meanP & 41.3\textcolor{blue}{(41.0)} & 73.2\textcolor{blue}{(73.0)} & 84.4\textcolor{blue}{(84.2)} & \textbf{2.0}\textcolor{blue}{(2.0)} & \textbf{7.6}\textcolor{blue}{(7.7)} & 198.9\textcolor{blue}{(198.2)} & \textbf{40.8}\textcolor{blue}{(40.8)} & 67.8\textcolor{blue}{(67.8)} & 76.4\textcolor{blue}{(77.1)} & \textbf{2.0}\textcolor{blue}{(2.0)} & 21.8\textcolor{blue}{(21.5)} & 185.0\textcolor{blue}{(185.7)} \\
			seqLSTM + expansion excitation + meanP & \textbf{41.7}\textcolor{blue}{(41.0)} & \textbf{73.8}\textcolor{blue}{(73.0)} & \textbf{84.8}\textcolor{blue}{(84.2)} & \textbf{2.0}\textcolor{blue}{(2.0)} & \textbf{7.6}\textcolor{blue}{(7.7)} & \textbf{200.3}\textcolor{blue}{(198.2)} & 39.9\textcolor{blue}{(40.8)} & \textbf{68.1}\textcolor{blue}{(67.8)} & \textbf{77.2}\textcolor{blue}{(77.1)} & \textbf{2.0}\textcolor{blue}{(2.0)} & \textbf{20.9}\textcolor{blue}{(21.5)} & \textbf{185.2}\textcolor{blue}{(185.7)} \\ 
			\hline
			seqLSTM + squeeze aggregation & 39.5\textcolor{blue}{(41.0)} & 70.1\textcolor{blue}{(73.0)} & 82.6\textcolor{blue}{(84.2)} & \textbf{2.0}\textcolor{blue}{(2.0)} & 11.7\textcolor{blue}{(7.7)} & 192.2\textcolor{blue}{(198.2)} & 37.3\textcolor{blue}{(40.8)} & 64.0\textcolor{blue}{(67.8)} & 74.3\textcolor{blue}{(77.1)} & 3.0\textcolor{blue}{(2.0)} & 30.2\textcolor{blue}{(21.5)} & 175.6\textcolor{blue}{(185.7)} \\
			seqLSTM + expansion aggregation & \textbf{42.6}\textcolor{blue}{(41.0)} & \textbf{74.5}\textcolor{blue}{(73.0)} & \textbf{85.3}\textcolor{blue}{(84.2)} & \textbf{2.0}\textcolor{blue}{(2.0)} & \textbf{8.0}\textcolor{blue}{(7.7)} & \textbf{202.4}\textcolor{blue}{(198.2)} & \textbf{38.8}\textcolor{blue}{(40.8)} & \textbf{68.3}\textcolor{blue}{(67.8)} & \textbf{75.8}\textcolor{blue}{(77.1)} & \textbf{2.0}\textcolor{blue}{(2.0)} & \textbf{22.0}\textcolor{blue}{(21.5)} & \textbf{182.9}\textcolor{blue}{(185.7)} \\  
			\hline
			seqLSTM + squeeze excitation + squeeze aggregation & 40.5\textcolor{blue}{(41.0)} & 72.2\textcolor{blue}{(73.0)} & 83.1\textcolor{blue}{(84.2)} & \textbf{2.0}\textcolor{blue}{(2.0)} & 11.4\textcolor{blue}{(7.7)} & 195.8\textcolor{blue}{(198.2)} & 37.7\textcolor{blue}{(40.8)} & 64.9\textcolor{blue}{(67.8)} & 75.4\textcolor{blue}{(77.1)} & \textbf{2.0}\textcolor{blue}{(2.0)} & 27.4\textcolor{blue}{(21.5)} & 178.0\textcolor{blue}{(185.7)} \\
			seqLSTM + expansion excitation + expansion aggregation & \textbf{42.6}\textcolor{blue}{(41.0)} & \textbf{74.2}\textcolor{blue}{(73.0)} & \textbf{85.3}\textcolor{blue}{(84.2)} & \textbf{2.0}\textcolor{blue}{(2.0)} & 8.2\textcolor{blue}{(7.7)} & \textbf{202.1}\textcolor{blue}{(198.2)} & \textbf{40.6}\textcolor{blue}{(40.8)} & \textbf{68.4}\textcolor{blue}{(67.8)} & \textbf{76.4}\textcolor{blue}{(77.1)} & \textbf{2.0}\textcolor{blue}{(2.0)} & \textbf{21.3}\textcolor{blue}{(21.5)} & \textbf{185.4}\textcolor{blue}{(185.7)} \\ 
			seqLSTM + squeeze excitation + expansion aggregation & 42.3\textcolor{blue}{(41.0)} & 74.0\textcolor{blue}{(73.0)} & 84.9\textcolor{blue}{(84.2)} & \textbf{2.0}\textcolor{blue}{(2.0)} & \textbf{8.1}\textcolor{blue}{(7.7)} & 201.2\textcolor{blue}{(198.2)} & 39.6\textcolor{blue}{(40.8)} & 67.3\textcolor{blue}{(67.8)} & 76.1\textcolor{blue}{(77.1)} & \textbf{2.0}\textcolor{blue}{(2.0)} & 21.5\textcolor{blue}{(21.5)} & 183.0\textcolor{blue}{(185.7)} \\ 
			seqLSTM + expansion excitation + squeeze aggregation & 41.9\textcolor{blue}{(41.0)} & 73.7\textcolor{blue}{(73.0)} & 84.5\textcolor{blue}{(84.2)} & \textbf{2.0}\textcolor{blue}{(2.0)} & 8.8\textcolor{blue}{(7.7)} & 200.1\textcolor{blue}{(198.2)} & 39.0\textcolor{blue}{(40.8)} & 66.2\textcolor{blue}{(67.8)} & 74.7\textcolor{blue}{(77.1)} & \textbf{2.0}\textcolor{blue}{(2.0)} & 24.2\textcolor{blue}{(21.5)} & 179.9\textcolor{blue}{(185.7)} \\ 
			\hline
		\end{tabular}
	}
\end{table*}

\begin{table*}[ht]
	\caption{T2V retrieval results with different frame aggregation methods in the seqTransf type on the ActivityNet and DiDeMo datasets. The number in blue is the retrieval result obtained by our sequential Transformer Encoder re-training.}
	\centering
	\renewcommand{\arraystretch}{1.1} 
	\label{table: seqTransf Type}
	\resizebox{\textwidth}{!}{
		\begin{tabular}{l|cccccc|cccccc}
			\hline
			\multirow{2}*{Methods} &  \multicolumn{6}{c|}{ActivityNet Retrieval} & \multicolumn{6}{c}{DiDeMo Retrieval} \\ 
			\cline{2-7} \cline{8-13} 
			& R@1$\uparrow$  & R@5$\uparrow$  & R@10$\uparrow$ & MdR$\downarrow$ &  MnR$\downarrow$ & RSum$\uparrow$
			& R@1$\uparrow$ & R@5$\uparrow$  & R@10$\uparrow$ & MdR$\downarrow$ &  MnR$\downarrow$ &
			RSum$\uparrow$ \\
			\hline
			seqTransf + squeeze excitation + meanP & \textbf{42.7}\textcolor{blue}{(42.2)} & 73.7\textcolor{blue}{(74.1)} & \textbf{85.2}\textcolor{blue}{(85.2)} & \textbf{2.0}\textcolor{blue}{(2.0)} & 7.2\textcolor{blue}{(7.2)} & 201.6\textcolor{blue}{(201.5)} & 39.8\textcolor{blue}{(40.6)} & 66.9\textcolor{blue}{(66.3)} & \textbf{76.2}\textcolor{blue}{(76.1)} & \textbf{2.0}\textcolor{blue}{(2.0)} & \textbf{18.7}\textcolor{blue}{(19.8)} & 182.9\textcolor{blue}{(183.0)} \\
			seqTransf + expansion excitation + meanP & 42.6\textcolor{blue}{(42.2)} & \textbf{74.4}\textcolor{blue}{(74.1)} & 85.1\textcolor{blue}{(85.2)} & \textbf{2.0}\textcolor{blue}{(2.0)} & \textbf{7.1}\textcolor{blue}{(7.2)} & \textbf{202.1}\textcolor{blue}{(201.5)} & \textbf{41.4}\textcolor{blue}{(40.6)} & \textbf{67.0}\textcolor{blue}{(66.3)} & 76.0\textcolor{blue}{(77.1)} & \textbf{2.0}\textcolor{blue}{(2.0)} & 19.7\textcolor{blue}{(19.8)} & \textbf{184.4}\textcolor{blue}{(183.0)} \\ 
			\hline
			seqTransf + squeeze aggregation & 41.4\textcolor{blue}{(44.2)} & 73.1\textcolor{blue}{(74.1)} & 84.7\textcolor{blue}{(85.2)} & \textbf{2.0}\textcolor{blue}{(2.0)} & 8.3\textcolor{blue}{(7.2)} & 199.2\textcolor{blue}{(201.5)} & 37.6\textcolor{blue}{(40.6)} & 64.5\textcolor{blue}{(66.3)} & 74.9\textcolor{blue}{(76.1)} & \textbf{2.0}\textcolor{blue}{(2.0)} & 22.8\textcolor{blue}{(19.8)} & 177.0\textcolor{blue}{(183.0)} \\
			seqTransf + expansion aggregation & \textbf{44.1}\textcolor{blue}{(42.2)} & \textbf{75.1}\textcolor{blue}{(74.1)} & \textbf{86.2}\textcolor{blue}{(85.2)} & \textbf{2.0}\textcolor{blue}{(2.0)} & \textbf{7.1}\textcolor{blue}{(7.2)} & \textbf{205.4}\textcolor{blue}{(201.5)} & \textbf{39.2}\textcolor{blue}{(40.6)} & \textbf{65.7}\textcolor{blue}{(66.3)} & \textbf{75.7}\textcolor{blue}{(76.1)} & \textbf{2.0}\textcolor{blue}{(2.0)} & \textbf{20.1}\textcolor{blue}{(19.8)} & \textbf{180.6}\textcolor{blue}{(183.0)} \\  
			\hline
			seqTransf + squeeze excitation + squeeze aggregation & 42.3\textcolor{blue}{(42.2)} & 73.8\textcolor{blue}{(74.1)} & 84.9\textcolor{blue}{(85.2)} & \textbf{2.0}\textcolor{blue}{(2.0)} & 8.3\textcolor{blue}{(7.2)} & 201.0\textcolor{blue}{(201.5)} & 37.8\textcolor{blue}{(40.6)} & 64.6\textcolor{blue}{(66.3)} & 75.0\textcolor{blue}{(76.1)} & 3.0\textcolor{blue}{(2.0)} & 22.8\textcolor{blue}{(19.8)} & 177.4\textcolor{blue}{(183.0)} \\
			seqTransf + expansion excitation + expansion aggregation & \textbf{43.9}\textcolor{blue}{(42.2)} & \textbf{75.4}\textcolor{blue}{(74.1)} & 85.9\textcolor{blue}{(85.2)} & \textbf{2.0}\textcolor{blue}{(2.0)} & \textbf{7.2}\textcolor{blue}{(7.2)} & \textbf{205.2}\textcolor{blue}{(201.5)} & \textbf{39.5}\textcolor{blue}{(40.6)} & \textbf{66.1}\textcolor{blue}{(66.3)} & \textbf{77.0}\textcolor{blue}{(76.1)} & \textbf{2.0}\textcolor{blue}{(2.0)} & \textbf{19.4}\textcolor{blue}{(19.8)} & \textbf{182.6}\textcolor{blue}{(183.0)} \\ 
			seqTransf + squeeze excitation + expansion aggregation & \textbf{43.9}\textcolor{blue}{(42.2)} & \textbf{75.4}\textcolor{blue}{(74.1)} & 85.6\textcolor{blue}{(85.2)} & \textbf{2.0}\textcolor{blue}{(2.0)} & \textbf{7.2}\textcolor{blue}{(7.7)} & 204.9\textcolor{blue}{(201.5)} & 39.0\textcolor{blue}{(40.6)} & 65.6\textcolor{blue}{(66.3)} & 76.3\textcolor{blue}{(76.1)} & \textbf{2.0}\textcolor{blue}{(2.0)} & 20.7\textcolor{blue}{(19.8)} & 180.9\textcolor{blue}{(183.0)} \\ 
			seqTransf + expansion excitation + squeeze aggregation & 43.2\textcolor{blue}{(42.2)} & 74.2\textcolor{blue}{(74.1)} & \textbf{86.0}\textcolor{blue}{(85.2)} & \textbf{2.0}\textcolor{blue}{(2.0)} & 7.7\textcolor{blue}{(7.2)} & 203.4\textcolor{blue}{(201.5)} & 38.7\textcolor{blue}{(40.6)} & 65.4\textcolor{blue}{(66.3)} & 74.7\textcolor{blue}{(76.1)} & \textbf{2.0}\textcolor{blue}{(2.0)} & 20.4\textcolor{blue}{(19.8)} & 178.8\textcolor{blue}{(183.0)} \\ 
			\hline
		\end{tabular}
	}
\end{table*}

\end{document}